\documentclass[aps,prd,showpacs,superscriptaddress,nofootinbib]{revtex4-2} 
\usepackage{amsmath,amssymb,bm,graphicx}
\usepackage{xcolor}
\usepackage[hidelinks]{hyperref}
\usepackage{siunitx}
\usepackage{placeins}

\usepackage{orcidlink}
\DeclareUnicodeCharacter{2212}{$-$}

\newcommand{\oo}{w_0}
\newcommand{\wa}{w_a}

\newenvironment{revblock}{}{}

\begin{document}

\title{Controlled Ensemble Null Tests of Dynamical Dark Energy Hints: Single-Probe Recovery of $\Lambda$CDM with BAO, Compressed-CMB, and SNe Mocks}


\author{Seokcheon Lee \orcidlink{0000-0003-0861-1300}}
\email{skylee@skku.edu}
\affiliation{Department of Physics, Institute of Basic Science, Sungkyunkwan University, Suwon 16419, Korea}
\date{\today}


\begin{abstract}
\begin{revblock}
Recent combined cosmological analyses have reported mild apparent departures from the cosmological-constant equation of state. Such departures are difficult to interpret because BAO, compressed CMB distance priors, and Type Ia supernovae constrain different projections of the expansion history and carry different calibration degeneracies. We perform controlled ensemble null tests using observable-level Gaussian BAO-like, CMB-like, and supernova-like mock datasets generated from a common fiducial flat $\Lambda$CDM cosmology. The aim is to determine whether each constituent single-probe likelihood block can produce an apparent DDE-like marginalized-posterior displacement, or spuriously exclude the fiducial model, through its own degeneracy structure and calibration freedom. The analysis adopts probe-adapted parameterizations for the BAO scale, the CMB physical matter density, and the supernova normalization degeneracy. With explicit convergence diagnostics, percentile-based posterior coverage tests, and realization-level pull statistics, none of the three single-probe analyses spuriously excludes the fiducial input at the 95 per cent level. All completed BAO and SNe chains enter their final ensembles after adaptive extension of the originally rejected runs; for compressed CMB, a mock-data-space audit finds no detectable selection dependence within the sensitivity of the 208-mock ensemble. The compressed CMB-only likelihood nevertheless exhibits stable displaced marginalized medians together with broad late-time posterior intervals, illustrating that posterior location and model exclusion are distinct diagnostics. These results show that a displaced marginalized posterior in a single realization is not, by itself, evidence for dynamical dark energy. Such shifts must be interpreted through mock-ensemble recovery, convergence diagnostics, and the probe-dependent geometry of the likelihood.
\end{revblock}
\end{abstract}

\maketitle

\tableofcontents


\section{Introduction}
\label{sec:intro}
\begin{revblock}
Recent multi-probe analyses have reported apparent deviations from the cosmological-constant equation of state (EOS), $(w_0,w_a)=(-1,0)$, often interpreted as hints of dynamical dark energy (DDE)~\cite{DESI:2024mwx,DESI:2025zgx,DESI:2025fii,DES:2025tir,Ishak:2025cay}. These claims have stimulated a broad reassessment of how robustly current data distinguish genuine time variation in dark energy from probe-level inconsistencies, calibration systematics, or limitations of the adopted parametrization. Supernova analyses have emphasized the role of absolute calibration, low-redshift systematics, and the inferred matter density in shaping apparent DDE preferences~\cite{Brout:2022vxf,DES:2024fdw,Efstathiou:2024xcq,Colgain:2024mtg,Huang:2025som,Lee:2025kbn}. Related studies have also shown that the interpretation of evolving-dark-energy evidence depends sensitively on the statistical framework, prior choices, and the consistency of the datasets being combined~\cite{Smith:2022hwi,Handley:2019wlz,DiValentino:2021izs,Lee:2025ysg,Ong:2025utx}.

A complementary issue is that geometric observables do not measure the DE EOS directly.  Luminosity and angular-diameter distances constrain integrals of the expansion history, while BAO measurements constrain distance ratios involving the sound horizon and compressed CMB distance priors constrain high-redshift geometric combinations. This connects the present problem to the broader question of cosmological consistency: distance and expansion observables test restricted combinations of the underlying cosmological functions rather than independently determining all of them~\cite{Clarkson:2010uz,Heavens:2017hkr,Efstathiou:2019mdh}. Several recent analyses have made this limitation more explicit by showing that distance-based constraints on $w(z)$ are intrinsically smoothed by integral kernels and are therefore sensitive mainly to a small number of broad modes~\cite{Huterer:2004ch,Shlivko:2024llw,Giare:2025pzu,Lee:2025jrr,Lee:2026sta}. This motivates a controlled null test in which apparent DDE shifts are calibrated against mock ensembles rather than inferred from a single realization.

The CPL parametrization remains the standard two-parameter description of late-time DE evolution~\cite{Chevallier:2000qy,Linder:2002et,Albrecht:2006um}. However, the same flexibility that makes CPL useful also allows partially inconsistent or weakly constrained datasets to move along elongated directions in the $(w_0,w_a)$ plane. Such degeneracy directions have appeared repeatedly in observational DE analyses and in studies of whether current data really require extensions beyond $\Lambda$CDM~\cite{Wang:2017lai,DES:2024xij,Zhang:2018glx,Giare:2024gpk,Notari:2024zmi,Steinhardt:2025znn,Afroz:2025iwo,Duangchan:2025uzj}. From this perspective, a visually displaced contour in the $(w_0,w_a)$ plane is not, by itself, a robust physical statement. It must be interpreted together with the likelihood geometry, the relevant calibration freedoms, and the degree of inter-probe consistency.

Different probes constrain different functions of the expansion history. Low-redshift supernovae constrain the relative shape of the luminosity-distance relation and are nearly insensitive to the absolute normalization without calibration information~\cite{Brout:2022vxf,DES:2024fdw,Huang:2025som}. BAO measurements constrain dimensionless distance ratios such as $D_M/r_d$ and $D_H/r_d$, and thus directly involve the sound-horizon scale~\cite{DESI:2024mwx,DESI:2025zgx,DESI:2025fii,Notari:2024zmi,Giare:2024gpk,Lee:2025jrk}. Compressed CMB distance priors constrain high-redshift geometric combinations through the shift parameter, the acoustic scale, and physical densities~\cite{Chen:2018dbv,Zhai:2019nad,Lemos:2023xhs,Planck:2018vyg}. Consequently, the degeneracy directions associated with BAO, CMB, and SNe are not aligned in the $(w_0,w_a)$ plane. Therefore, posterior medians from individual probes can move along broad degeneracy ridges even when the input cosmology is exactly $\Lambda$CDM.

The aim of this paper is to turn this qualitative degeneracy statement into a controlled ensemble null test. We generate mutually consistent BAO-like, CMB-like, and SNe mock ensembles from the same fiducial flat $\Lambda$CDM cosmology and ask whether the corresponding single-probe likelihood blocks recover the input model in posterior coverage, pull statistics, and convergence diagnostics. This ensemble formulation is essential because a single realization can display visually displaced marginal contours along a weakly constrained ridge even when the likelihood block is statistically well calibrated. It also complements controlled tension-injection studies, which show how explicit cross-probe mismatches in quantities such as $\Omega_{m0}$, $H_0$, $M$, or $r_d$ can propagate into apparent shifts in $w_0$, $w_a$, and the CPL parameters~\cite{Tang:2024lmo,DES:2025bxy,Chen:2025jnr,Lee:2025axp}.

The analysis adopts parameterizations that follow the dominant degeneracy directions of each probe. For BAO-only fits we sample $q_{\rm BAO}\equiv h r_d$, because BAO distances constrain $H_0 r_d$ rather than $H_0$ and $r_d$ independently. For CMB-only fits we sample $\omega_m=\Omega_{m0}h^2$ internally and transform back to $\Omega_{m0}$ for reporting. For SNe-only fits we sample $\Delta M_{\rm eff}=\Delta M-5\log_{10}(H_0/H_{0,\rm ref})$, with $H_{0,\rm ref}=70~{\rm km\,s^{-1}\,Mpc^{-1}}$, which follows the exact normalization degeneracy of an uncalibrated Hubble diagram. These choices avoid treating poorly constrained calibration directions as independently measured quantities.

The combined likelihoods considered in the final part of the analysis are products of these single-probe likelihood factors. They are discussed only to clarify how BAO, CMB, and SNe degeneracy directions are expected to intersect once the individually tested likelihood blocks are multiplied.

The target of the present study is the closure and posterior geometry of the constituent likelihood blocks rather than the frequentist calibration of their full product. Specifically, we ask whether an individual BAO-only, compressed CMB-only, or SNe-only likelihood block, generated from a common fiducial flat $\Lambda$CDM cosmology, can produce an apparent DDE-like marginalized-posterior displacement or spuriously exclude the fiducial model through its own degeneracy structure and calibration freedom.

This question is distinct from the ensemble validation of the full BAO+CMB+SNe product likelihood. Recovery of the constituent single-probe blocks does not mathematically guarantee correct coverage of their product, because the intersection of non-aligned likelihood ridges can amplify realization-dependent offsets. A matched combined-probe ensemble would require additional choices concerning realization matching, cross-probe covariance, joint chain-quality selection, and the definition of multidimensional combined coverage. These choices define a separate statistical experiment rather than a direct extension of the constituent-likelihood closure tests performed here. Accordingly, the statistical null-test claim of this work is restricted to the ensemble recovery and posterior geometry of the individual likelihood blocks.

The rest of the paper is organized as follows. Section~\ref{sec:theory} reviews the CPL parametrization and the probe-dependent likelihood geometry used for interpretation. Section~\ref{sec:methodology} describes the mock construction, parameterizations, convergence diagnostics, and coverage statistics. Section~\ref{sec:mock_likelihoods} defines the BAO, CMB, and SNe mock likelihoods. Section~\ref{sec:results} presents the ensemble null-test results. Section~\ref{sec:conclusions} summarizes the implications for interpreting apparent DDE hints.
\end{revblock}

\section{CPL Degeneracy and Probe-Dependent Likelihood Geometry}
\label{sec:theory}

In this section we summarize the theoretical ingredients used to interpret the three constituent-likelihood null tests. We first review the origin of the $(\oo,\wa)$ degeneracy in the Chevallier--Polarski--Linder (CPL) parametrization and explain why BAO, compressed CMB distance priors, and SNe trace different directions in this parameter space. We then discuss a Fisher-based shift relation and the geometric compensation among the probes. These tools are used only to interpret the posterior geometry of the individual likelihood blocks; the frequentist calibration of their full product is outside the scope of the present analysis.

\subsection{CPL degeneracy and its geometric origin}
\label{subsec:cpl_geometry}

The CPL parametrization~\cite{Chevallier:2000qy,Linder:2002et},
\begin{equation}
    w(z)=\oo+\wa\frac{z}{1+z},
    \label{CPL}
\end{equation}
remains the standard two-parameter description of a time-dependent DE EOS. Its usefulness lies not only in its simple low-redshift form but also in the fact that different probes respond to different combinations of $(\oo,\wa)$.

For a flat background cosmology, the expansion history may be written as
\begin{equation}
   \frac{H^2(z)}{H_0^2}\equiv E^2(z)
   = \Omega_{r0}(1+z)^4+\Omega_{m0}(1+z)^3
   +(1-\Omega_{m0}-\Omega_{r0})(1+z)^{3(1+\oo+\wa)}
  e^{-\frac{3\wa z}{1+z}} \,,
   \label{Ez}
\end{equation}
where $\Omega_{r0}$ and $\Omega_{m0}$ are the present-day radiation and matter density fractions. Small compensating changes in $(\oo,\wa)$ can leave $E(z)$ nearly unchanged over a limited redshift range. This is the origin of the familiar elongated ridge in the $(\oo,\wa)$ plane. In practice, one often finds an approximate relation of the form~\cite{Lee:2025ysg}
\begin{equation}
    \delta \wa \simeq -A\,\delta \oo ,
\end{equation}
where the coefficient $A$ depends on the redshift range and the observable under consideration.

This is why different probes produce different degeneracy directions. SNe depend on the luminosity distance,
\begin{equation}
    D_L(z)=(1+z)\frac{c}{H_0}\int_0^z \frac{dz'}{E(z')},
\end{equation}
and therefore constrain the relative late-time expansion history. BAO constrain dimensionless distance combinations such as $D_M/r_d$ and $D_H/r_d$, and are sensitive to a different mixture of $\Omega_{m0}$, $w_0$, and $w_a$. Compressed CMB distance priors constrain the high-redshift geometry through quantities such as the shift parameter $R$ and the acoustic scale $\ell_A$. The result is that BAO, CMB, and SNe do not pull along the same direction in the $(\oo,\wa)$ plane~\cite{Lee:2026ets}.

\subsection{Bias equation under partial or inconsistent likelihoods}
\label{subsec:bias_equation}

When multiple probes are combined, shifts in the inferred parameters can arise even if the underlying cosmology is perfectly consistent with $\Lambda$CDM~\cite{Tang:2024lmo,DES:2025bxy,Chen:2025jnr,Lee:2025jrk}. The reason is not necessarily new physics. It can also come from incomplete likelihood information, implicit priors, calibration freedom, or simply from the fact that different probes are sensitive to different directions in parameter space.

In the Gaussian approximation, suppose probe $i$ has a small offset $\delta\boldsymbol{\theta}_i$ between its preferred location and the true fiducial point. Then the shift of the combined posterior mean is
\begin{equation}
    \Delta\boldsymbol{\theta} = F^{-1}\sum_i F_i\,\delta\boldsymbol{\theta}_i , \label{eq:bias_equation}
\end{equation}
where $F_i$ is the Fisher matrix of probe $i$ and
\begin{equation}
    F=\sum_i F_i
\end{equation}
is the Fisher matrix of the full combination. \begin{revblock}This expression is the standard linearized Fisher-bias relation used to propagate small probe-dependent offsets into parameter space~\cite{Tegmark:1996qs,Heavens:2017hkr,Lee:2026ets}. In the null tests considered here, effective posterior displacements can reflect realization-dependent fluctuations, nonlinear marginalization of broad likelihood ridges, prior-volume effects associated with the adopted sampling coordinates, and residual numerical sampling behavior. They do not arise from a departure of the mock-generating cosmology from $\Lambda$CDM. \end{revblock}

Its geometric meaning is straightforward. \begin{revblock} Each probe defines a local likelihood curvature around its preferred point, represented by an error ellipse or, in a weakly constrained subspace, an elongated ridge. The quantity $F_i\,\delta\boldsymbol{\theta}_i$ weights the probe-dependent offset according to the directions in which that probe has the greatest likelihood curvature, while $F^{-1}$ converts the sum of these weighted contributions into the net displacement of the combined posterior. When different probes have long, non-aligned degeneracy ridges, small relative displacements of their individual likelihoods can shift the location of their intersection appreciably. The effect is largest along directions that remain weakly constrained by the full combination, for which the corresponding eigenvalues of $F$ are small. \end{revblock}

In the present context, the relevant degeneracy structure is especially transparent. SNe constrain the redshift dependence of the relative luminosity distance, or equivalently the shape of the Hubble diagram, but do not determine the absolute distance scale once the supernova absolute magnitude is marginalized over. BAO constrain $D_M/r_d$ and $D_H/r_d$, so their parameter-space geometry reflects both the late-time expansion history and its degeneracy with the sound-horizon scale $r_d$. The compressed CMB distance priors constrain high-redshift geometric combinations through $(R,\ell_A,\omega_b)$ and therefore select directions in the CPL parameter space that differ from those constrained by BAO and SNe. Because these dominant directions are not aligned, an individual partial combination need not be centered exactly on $(\oo,\wa)=(-1,0)$ even when all mock data are generated from the same fiducial $\Lambda$CDM cosmology.

Therefore, the point is not only that an explicit inconsistency between data sets can shift the inferred EOS. Even in the absence of a genuine inconsistency, realization-dependent offsets can produce displaced joint contours because different probes constrain different combinations of the same parameters. \begin{revblock} Partial probe combinations are particularly susceptible because important degeneracy directions may remain weakly constrained. A reduced or compressed likelihood can introduce an additional effect only if it fails to preserve all of the information relevant to those directions. This is the basic mechanism examined by the mock null tests in this paper. \end{revblock}

\subsection{Geometric compensation in the CMB--BAO--SNe system}
\label{sec:CMB_DE_generacy}

A familiar example of multi-probe compensation arises from the different parameter combinations constrained by CMB and late-time distance observables. In a spatially flat geometry, the comoving angular-diameter distance to last scattering is
\begin{equation}
    D_M(z_*) = \frac{c}{H_0}\int_0^{z_*}\frac{dz'}{E(z')}.
\end{equation}
More directly, the CMB acoustic scale constrains the ratio $r_s(z_*)/D_M(z_*)$. Therefore, for a fixed early-time calibration and approximately fixed physical matter densities, a change in $H_0$ must be compensated by a corresponding change in the dimensionless expansion history.

Within the flat CPL parameterization considered here, the local CMB distance degeneracy typically correlates a lower $H_0$ with a larger $\Omega_{m0}$ and with compensating shifts toward larger $w_0$ and smaller $w_a$~\cite{Lee:2025ysg,Lee:2025jrk,Lee:2025kbn}. Part of this behavior may approximately follow a direction of nearly constant early-time CPL asymptote,
\begin{equation}
    \oo+\wa \simeq \mathrm{const.},  \label{oowa}
\end{equation}
although the actual distance-degeneracy direction is determined by the integrated expansion history and need not coincide exactly with this relation.

BAO measurements reduce part of this freedom by constraining $D_M/r_d$ and $D_H/r_d$ at intermediate redshifts. Thus, their geometry depends on both the late-time expansion history and its degeneracy with the sound-horizon scale $r_d$. SNe provide a complementary constraint on the redshift dependence of the relative luminosity distance, or equivalently on the shape of the Hubble diagram. They do not determine the absolute distance scale by themselves, but they rotate and narrow the allowed late-time region and thereby restrict the compensation that remains after combining BAO with CMB information alone~\cite{Lee:2026ets}.

Schematically, under the assumptions above, the relevant local CMB degeneracy may be represented as
\begin{equation}
    H_0 \downarrow \qquad \Rightarrow \qquad \Omega_{m0}\uparrow,\qquad \oo \uparrow,\qquad \wa \downarrow , \nonumber
\end{equation}
where the arrows describe a qualitative posterior direction rather than a universal parameter relation. The inclusion of SNe limits this motion by constraining the low-redshift relative-distance history more tightly. Table~\ref{tab:H0_wa_compensation} summarizes the complementary roles of the three probes. The CMB and BAO rows indicate the schematic local compensation directions discussed above, whereas the SNe row describes how the relative-distance information restricts the extent of this motion without independently fixing the absolute distance scale.

\begin{table}[t]
\centering
\caption{Qualitative roles of CMB, BAO, and SNe in the geometric compensation among $H_0$, $\Omega_{m0}$, $w_0$, and $w_a$. The arrows indicate local posterior trends under a fixed early-time calibration, rather than universal parameter relations.}
\vspace{0.5em}
\begin{tabular}{c|cccc}
\hline\hline
Probe 
& $H_0$ 
& $\Omega_{m0}$
& $w_0$
& $w_a$ \\
\hline
CMB
& $\downarrow$
& $\uparrow$
& $\uparrow$ (less negative)
& $\downarrow$ \\
BAO at fixed $r_d$
& $\uparrow$
& $\downarrow$
& $\downarrow$
& $\uparrow$ \\
SNe
& not fixed
& restricts drift
& tightens late-time value
& restricts evolution \\
\hline\hline
\end{tabular}
\label{tab:H0_wa_compensation}
\end{table}


\section{Mock Construction, Parameterization, and Recovery Diagnostics}
\label{sec:methodology}
\begin{revblock}
This section describes the numerical strategy used for the ensemble null tests. The analysis is based on mock ensembles rather than on a single realization, so that realization-to-realization scatter can be separated from systematic posterior displacement. The goal is to test whether each single-probe likelihood block retains the common fiducial $\Lambda$CDM input with the expected posterior coverage and pull behavior after applying explicit chain-quality cuts, rather than requiring the marginal posterior medians of underconstrained likelihoods to coincide with the fiducial point.

\subsection{Fiducial cosmology and mock ensembles}
\label{subsec:method_setup}

All mock datasets are generated around a common flat $\Lambda$CDM background,
\begin{equation}
    \Omega_{m0}=0.30,\qquad
    H_0=70.0~{\rm km\,s^{-1}\,Mpc^{-1}},\qquad
    w_0=-1.0,\qquad
    w_a=0.0 .
\end{equation}
For the compressed CMB likelihood, we additionally adopt
\begin{equation}
    \omega_b\equiv\Omega_bh^2=0.02237,
\end{equation}
with $T_{\rm CMB}=2.7255~{\rm K}$ and $N_{\rm eff}=3.046$ fixed in the calculation of the distance-prior theory vector.

For the SNe-only normalization coordinate, we define the dimensionful reference value
\begin{equation}
    H_{0,\rm ref}\equiv70~{\rm km\,s^{-1}\,Mpc^{-1}}.
\end{equation}

For the BAO-only mock ensemble, the directly relevant scale coordinate is specified by
\begin{equation}
    q_{\rm BAO}^{\rm fid}\equiv h r_d=103.0~{\rm Mpc}.
\end{equation}
Because the BAO observables used here depend on $h$ and $r_d$ only through this product, no separate value of $r_d$ is required for the BAO mock generation.

For each probe, we generate correlated Gaussian mock realizations using the corresponding covariance matrix and fiducial theory vector. The analysis focuses on the BAO-only, CMB-only, and SNe-only ensembles, because these likelihood blocks expose the relevant degeneracy directions and nuisance-parameter choices most directly.

\subsection{Degeneracy-respecting parameterizations}
\label{subsec:reparam}
The BAO-only analysis samples
\begin{equation}
\Theta_{\rm BAO}=(\Omega_{m0},w_0,w_a,q_{\rm BAO}),
\qquad
q_{\rm BAO}\equiv h r_d,
\end{equation}
rather than sampling $H_0$ and $r_d$ independently. This follows from the fact that BAO observables are $D_M/r_d$ and $D_H/r_d$, which are primarily sensitive to the product $H_0 r_d$, while retaining an additional dependence on the dimensionless expansion history through $\Omega_{m0}$, $w_0$, and $w_a$. This reparameterization removes the redundant long $H_0$--$r_d$ sampling direction and avoids interpreting $H_0$ and $r_d$ as independently measured by BAO alone.

For the CMB-only ensemble, the sampler uses the internal parameter vector
\begin{equation}
\Theta_{\rm CMB}^{\rm int}=(\omega_m,H_0,w_0,w_a,\omega_b),
\qquad
\omega_m\equiv \Omega_{m0}h^2,
\end{equation}
and transforms back to
\begin{equation}
\Omega_{m0}=\frac{\omega_m}{h^2}
\end{equation}
for reporting. This parameterization follows the physical matter-density combination to which the compressed CMB distance prior is most directly sensitive, while retaining $H_0$ as a reported late-time parameter.

For the SNe-only ensemble, we use the exact normalization-degenerate combination
\begin{equation}
\Delta M_{\rm eff}=\Delta M-5\log_{10}\left(\frac{H_0}{H_{0,\rm ref}}\right).
\end{equation}
Consequently, the sampler uses $\Omega_{m0},H_0,w_0,w_a,\Delta M_{\rm eff}$, while the physical nuisance parameter reported in tables is
\begin{equation}
\Delta M=\Delta M_{\rm eff}+5\log_{10}\left(\frac{H_0}{H_{0,\rm ref}}\right).
\end{equation}
This is an exact reparameterization of the SNe-only likelihood and makes explicit that uncalibrated supernovae constrain relative luminosity distances rather than an independent absolute \(H_0\).

The notation $q_{\rm BAO}$ is used only for the BAO scale combination because it is a probe-specific sampling coordinate rather than a new physical observable. The analogous degeneracy-respecting coordinates for the other likelihood blocks already have standard physical interpretations: for the CMB distance-prior likelihood the natural coordinate is the physical matter density $\omega_m=\Omega_{m0}h^2$, while for the SNe-only likelihood it is the normalization-degenerate combination $\Delta M_{\rm eff}=\Delta M-5\log_{10}(H_0/H_{0,\rm ref})$.

\subsection{MCMC sampling and convergence diagnostics}
\label{subsec:mcmc}

All posteriors are sampled with the affine-invariant ensemble sampler \texttt{emcee}~\cite{Foreman-Mackey:2012any}. For every mock realization, we compute the maximum integrated autocorrelation time $\tau_{\max}$, the ratio $N_{\rm prod}/\tau_{\max}$, the mean acceptance fraction, and a walker-based split-$\hat R_{\max}$ statistic, where the maximum is taken over the sampled parameters. In this implementation, split segments of the interacting \texttt{emcee} walkers are treated as pseudo-independent chains. We therefore use split-$\hat R$ as a conservative internal heuristic rather than as a formally calibrated Gelman--Rubin statistic for independent chains. Chains are not accepted merely because the sampler terminates; instead, they are classified using explicit quality cuts.

For the BAO-only and SNe-only ensembles, we require
\begin{equation}
    N_{\rm prod}/\tau_{\max}>40,
    \qquad
    \hat R_{\max}<1.08 .
\end{equation}
The initial fixed-length analysis left 60 of 192 completed BAO runs and 13 of 116 completed SNe runs outside these cuts. We reran those realizations adaptively, using the same stored mock vectors, likelihoods, priors, parameterizations, and initialization procedure, until both criteria were satisfied. All 60 BAO and all 13 SNe reruns ultimately passed, so the final BAO and SNe recovery ensembles contain every completed run and are not conditioned on a convergence-based subset. For both BAO and SNe, the ensemble posterior summaries change only modestly when the originally rejected chains are carried to convergence and included, while the 95 per cent recovery conclusions remain unchanged.

The CMB-only compressed likelihood is more weakly identified because the three observables $(R,\ell_A,\omega_b)$ constrain a five-dimensional sampled parameter space. The resulting broad degeneracy directions make the autocorrelation and split-chain diagnostics more difficult to satisfy uniformly. Therefore, we adopt the slightly relaxed baseline criteria
\begin{equation}
    N_{\rm prod}/\tau_{\max}>35,
    \qquad
    \hat R_{\max}<1.09 ,
\end{equation}
for the main CMB-only recovery analysis. As a robustness check, we repeat the CMB analysis using the stricter common criteria
\begin{equation}
    N_{\rm prod}/\tau_{\max}>40,
    \qquad
    \hat R_{\max}<1.08 .
\end{equation}
For the CMB-only baseline, 101 of 208 fixed-length runs satisfy the primary criteria and 107 do not. Representative extensions to 100,000 production steps reached $N_{\rm prod}/\tau_{\max}=145.0$ and 117.9 while retaining $\hat R_{\max}=1.100$ and 1.153, respectively. The limiting parameters were $\omega_b$ and transformed $\Omega_{m0}$, not $w_0$ or $w_a$. The walkers occupied one connected CMB degeneracy ridge, and global first-half versus second-half ensemble means agreed to within about $0.035$ pooled posterior standard deviations for every parameter. Therefore, the persistent walker-based split-$\hat R$ reflects long-lived walker identity and ranking along that ridge rather than evidence for disconnected modes. The adaptive BAO and SNe reruns show explicitly that the original fixed-length quality selection does not materially alter the recovery conclusions once the initially rejected chains are carried to convergence. For the compressed-CMB likelihood, however, the long-chain pilots show that enforcing the same walker-based split-$\hat{R}$ threshold by brute-force extension can require very large additional sampling effort without clear statistical gain. This motivates treating the fixed-length quality cut as a practical computational filter whose potential selection effect must be validated directly in mock-data space. We therefore audit whether the baseline convergence classification depends on the mock-data fluctuations in Sec.~\ref{subsec:cmb_selection_audit} and Appendix~\ref{app:cmb_selection_audit}.

Accordingly, the final BAO and SNe coverages use all completed, converged runs. The reported CMB recovery statistics remain conditional on the 101-run baseline selection, with the 73-run stricter selection used only as a threshold robustness diagnostic. Within the statistical sensitivity of the full 208-mock audit, however, the baseline classification is not detectably associated with the amplitude or direction of the CMB mock fluctuations relevant to the cosmological inference.

\subsection{Ensemble recovery statistics}
\label{subsec:ensemble_stats}
For each accepted mock realization and each parameter \(\theta\), we report the posterior median and percentile intervals rather than assuming Gaussian symmetry. The 68 per cent coverage indicator tests direct inclusion of the fiducial value in $[\theta_{16},\theta_{84}]$, and the 95 per cent indicator tests direct inclusion in $[\theta_{2.5},\theta_{97.5}]$. No Gaussian or twice-asymmetric-$\sigma$ approximation is used for either coverage. We also compute the pull
\begin{equation}
p_\theta=\frac{\theta_{50}-\theta_{\rm fid}}{\sigma_{68}(\theta)},
\qquad
\sigma_{68}(\theta)=\frac{\theta_{84}-\theta_{16}}{2}.
\end{equation}
For non-Gaussian posteriors, this pull is only a compact diagnostic, not a replacement for the percentile intervals. We report both the 68 and 95 per cent marginal coverages, while treating the 95 per cent coverage and the joint rectangular coverage in the $(w_0,w_a)$ sector and in the full reported parameter set as the primary recovery diagnostics.
\end{revblock}


\section{Mock Likelihoods: DESI-like BAO, Planck-like CMB, and Pantheon-like SNe}
\label{sec:mock_likelihoods}
\begin{revblock}
We now summarize the likelihood blocks used in the ensemble validation. All mock data vectors are generated from the same fiducial flat $\Lambda$CDM cosmology. Thus, any apparent posterior displacement must arise from realization noise, likelihood geometry, prior-volume effects associated with the adopted sampling coordinates, or residual numerical sampling effects, rather than from an input inconsistency among the probes. 

For all probes, correlated Gaussian mock realizations are generated using the same convention. For each probe $X\in\{{\rm BAO},{\rm CMB},{\rm SN}\}$, we write
\begin{equation}
\mathbf d_X^{\rm mock} = \mathbf d_X^{\rm fid} + L_X\mathbf g_X, \qquad C_X=L_XL_X^T .
\label{eq:generic_mock_generation}
\end{equation}
Here $C_X$ is the fixed covariance matrix of probe $X$, $L_X$ is its Cholesky factor, and $\mathbf g_X$ is a vector of independent standard normal variates,
\begin{equation}
\mathbf g_X = (g_{X,1},\ldots,g_{X,n_X})^T, \qquad g_{X,\alpha}\sim{\cal N}(0,1), \qquad \langle g_{X,\alpha}g_{X,\beta}\rangle=\delta_{\alpha\beta}.
\label{eq:standard_normal_vector}
\end{equation}
The dimension $n_X$ is the length of the corresponding data vector. Thus, the Cholesky factors are probe dependent: $L_{\rm BAO}$ is built from the BAO covariance matrix, $L_{\rm CMB}$ from the CMB distance-prior covariance matrix, and $L_{\rm SN}$ from the full SNe covariance matrix~\cite{Tegmark:1996qs,Morrison:2013tqa,Wadekar:2020hax}.

For a model prediction $\mathbf d_X(\boldsymbol{\theta})$, the corresponding likelihood is
\begin{equation}
    -2\ln{\cal L}_X = \left[ \mathbf d_X^{\rm mock}-\mathbf d_X(\boldsymbol{\theta}) \right]^T C_X^{-1} \left[ \mathbf d_X^{\rm mock}-\mathbf d_X(\boldsymbol{\theta}) \right] +{\rm const.},                                                   
    \label{eq:generic_gaussian_likelihood}
\end{equation}
where the covariance matrices are treated as fixed and parameter independent.

The mocks used here are observable-level Gaussian realizations rather than field-level or catalogue-level simulations. They preserve the fiducial mean vectors and the full covariance structure adopted for each likelihood block, but they do not reproduce the upstream survey processes from which those observables were obtained. In particular, they do not explicitly model nonlinear structure formation, survey selection, reconstruction, calibration, or data-reduction systematics beyond the information already encoded in the adopted mean vectors and covariance matrices. Therefore, the resulting tests should be interpreted as controlled likelihood-level closure tests, rather than as end-to-end validations of the DESI, Planck, or Pantheon+ analysis pipelines.

This controlled construction is nevertheless appropriate for the present purpose because it isolates realization scatter, likelihood and posterior geometry, prior-volume effects associated with the adopted sampling coordinates, and residual numerical sampling behavior, while keeping the underlying cosmology exactly fixed.

\subsection{DESI-like BAO: \texorpdfstring{$D_M/r_d$}{DM/rd} and \texorpdfstring{$D_H/r_d$}{DH/rd}}

The BAO mock vector is built from DESI-like redshift sampling and covariance matrices~\cite{DESI:2024mwx,DESI:2025zgx}. For each redshift bin $z_i$, we compute
\begin{equation}
D_M(z_i)=\frac{c}{H_0}\int_0^{z_i}\frac{dz'}{E(z')}, \qquad D_H(z_i)=\frac{c}{H_0} \frac{1}{E(z_i)} .
\label{eq:bao_distances}
\end{equation}
Within the flat CPL parameterization used here, in which the dimensionless expansion history $E(z)$ is independent of the overall Hubble scale, the BAO ratios depend on $H_0$ and $r_d$ exactly through
the product
\begin{equation}
    q_{\rm BAO}\equiv h r_d \quad [{\rm Mpc}]  \label{eq:qbao_definition}.
\end{equation}
In the BAO-only analysis, however, the observables are the dimensionless ratios $D_M/r_d$ and $D_H/r_d$.  Since $H_0=100h\,{\rm km\,s^{-1}\,Mpc^{-1}}$, these ratios depend on $H_0$ and $r_d$ only through their product. Explicitly,
\begin{equation}
\frac{D_M(z)}{r_d} = \frac{c}{100\,q_{\rm BAO}} \int_0^z\frac{dz'}{E(z')}, \qquad
\frac{D_H(z)}{r_d} = \frac{c}{100\,q_{\rm BAO}}\frac{1}{E(z)} . \label{eq:bao_qbao_dependence}
\end{equation}
Thus, $q_{\rm BAO}$ is not an additional physical observable, but a degeneracy-respecting sampling coordinate for the BAO-only likelihood. For the fiducial model used in this work, we set $q_{\rm BAO}^{\rm fid}=103.0~{\rm Mpc}$ directly. No separate fiducial value of $r_d$ is required for the BAO-only mock generation.

The BAO data vector is the ordered collection of all DESI-like distance ratio entries included in the BAO covariance matrix. Schematically, for redshift bins in which both transverse and radial BAO information is used, it may be written as
\begin{equation}
\mathbf d_{\rm BAO} = \left( \frac{D_M(z_1)}{r_d},\ldots,\frac{D_M(z_N)}{r_d}, \frac{D_H(z_1)}{r_d},\ldots,\frac{D_H(z_N)}{r_d} \right)^T ,
\label{eq:bao_data_vector}
\end{equation}
with the actual ordering following the DESI-like covariance matrix. The BAO mock realization is generated as
\begin{equation}
\mathbf d_{\rm BAO}^{\rm mock} = \mathbf d_{\rm BAO}^{\rm fid} + L_{\rm BAO}\mathbf g_{\rm BAO}, \qquad C_{\rm BAO}=L_{\rm BAO}L_{\rm BAO}^T .
\label{eq:bao_mock_generation}
\end{equation}
Here $\mathbf g_{\rm BAO}$ has the same dimension as $\mathbf d_{\rm BAO}$.

Thus, we sample $(\Omega_{m0},w_0,w_a,q_{\rm BAO})$ in the BAO-only ensemble, rather than sampling $H_0$ and $r_d$ separately. This avoids treating $H_0$ and $r_d$ as independently measured quantities in a likelihood that primarily constrains their product.

\subsection{Planck-like CMB distance priors}

For the CMB, we use a compressed Gaussian likelihood based on
\begin{equation}
\mathbf v_{\rm CMB} = (R,\ell_A,\omega_b)^T,
\label{eq:cmb_data_vector}
\end{equation}
where
\begin{equation}
R=\sqrt{\Omega_{m0}}\frac{H_0D_M(z_*)}{c}, \qquad \ell_A=\pi\frac{D_M(z_*)}{r_s(z_*)}.
\label{eq:cmb_distance_priors}
\end{equation}
The CMB mock realization is generated as
\begin{equation}
\mathbf v_{\rm CMB}^{\rm mock} = \mathbf v_{\rm CMB}^{\rm fid} + L_{\rm CMB}\mathbf g_{\rm CMB}, \qquad C_{\rm CMB}=L_{\rm CMB}L_{\rm CMB}^T .
\label{eq:cmb_mock_generation}
\end{equation}
Here
\begin{equation}
\mathbf g_{\rm CMB} = (g_R,g_{\ell_A},g_{\omega_b})^T
\end{equation}
is a three-dimensional vector of independent standard normal variates, and $L_{\rm CMB}$ is the Cholesky factor of the Planck-like distance-prior covariance matrix~\cite{Chen:2018dbv,Zhai:2019nad,Planck:2018vyg,Lemos:2023xhs}.

The CMB-only sampler uses $\omega_m=\Omega_{m0}h^2$ internally. This avoids treating $\Omega_{m0}$ and $H_0$ as the most natural independent variables for the high-redshift compressed CMB likelihood.

\subsection{Pantheon-like supernovae with full covariance}

The supernova mock follows the statistical structure of the Pantheon+SH0ES compilation, including the full statistical and systematic covariance matrix~\cite{Brout:2022vxf}. For each supernova redshift $z_i$, the fiducial distance modulus is
\begin{equation}
\mu_{\rm fid}(z_i) = 5\log_{10}D_L(z_i)+25+\Delta M, \label{eq:sne_distance_modulus}
\end{equation}
where $D_L$ is measured in Mpc.  The corresponding SNe data vector is
\begin{equation}
\boldsymbol\mu_{\rm fid} = \left( \mu_{\rm fid}(z_1),\ldots,\mu_{\rm fid}(z_{N_{\rm SN}}) \right)^T .
\label{eq:sne_data_vector}
\end{equation}
A correlated Gaussian realization is generated as
\begin{equation}
\boldsymbol\mu_{\rm mock} = \boldsymbol\mu_{\rm fid} + L_{\rm SN}\mathbf g_{\rm SN}, \qquad C_{\rm SN}=L_{\rm SN}L_{\rm SN}^T .
\label{eq:sne_mock_generation}
\end{equation}
Here
\begin{equation}
\mathbf g_{\rm SN} = (g_1,\ldots,g_{N_{\rm SN}})^T
\end{equation}
is a vector of independent standard normal variates, and $L_{\rm SN}$ is the Cholesky factor of the full SNe covariance matrix.

The SNe-only likelihood is evaluated using the full covariance through the same covariance matrix $C_{\rm SN}$.  In the inference we sample $\Delta M_{\rm eff}$, which follows the exact $H_0$--$\Delta M$ normalization degeneracy, and transform back to $\Delta M$ only for reporting.

\subsection{Scope of the combined-probe discussion}

For statistically independent probe vectors, a formal combined likelihood would be the product of the constituent single-probe factors. No matched combined-probe ensemble is analyzed in the present work. The product likelihood is mentioned only to clarify how non-aligned single-probe degeneracy directions would be expected to intersect; its coverage and calibration are outside the scope of the present null tests.
\end{revblock}


\section{Ensemble Recovery of the Fiducial \texorpdfstring{$\Lambda$CDM}{LambdaCDM} Model}
\label{sec:results}
\begin{revblock}
This section presents the ensemble null-test results. The main question is whether each single-probe likelihood block artificially excludes the fiducial $\Lambda$CDM input when analyzed with the same covariance structure and parameterization used in the main inference. Therefore, we focus on posterior coverage, pull statistics, and explicit chain-quality cuts.  Combined-probe implications are discussed only geometrically and are not treated as independent ensemble-recovery results.

\subsection{BAO-only ensemble recovery}
\label{subsec:bao_ensemble}

The BAO-only analysis samples the parameter vector
\begin{equation}
    \boldsymbol{\theta}_{\rm BAO}
    =
    (\Omega_{m0},w_0,w_a,q_{\rm BAO}),
    \qquad
    q_{\rm BAO}\equiv h r_d.
\end{equation}
This choice follows directly from the structure of the BAO observables. The anisotropic BAO measurements constrain combinations such as $D_M(z)/r_d$ and $D_H(z)/r_d$, and therefore depend on the absolute distance scale primarily through $H_0r_d$, or equivalently $h r_d$. Sampling the directly constrained combination $q_{\rm BAO}$ removes the otherwise redundant long direction in the $(H_0,r_d)$ plane and provides a more efficient likelihood-level null test.

Each BAO mock realization is a correlated draw of the full BAO observable vector,
\begin{equation}
    {\bf d}_{\rm BAO}^{(k)}
    \sim
    {\cal N}
    \left(
        {\bf d}_{\rm BAO}^{\rm fid},
        {\bf C}_{\rm BAO}
    \right),
\end{equation}
where ${\bf d}_{\rm BAO}^{\rm fid}$ is evaluated at
\begin{equation}
    \Omega_{m0}=0.30,
    \qquad
    w_0=-1,
    \qquad
    w_a=0,
    \qquad
    q_{\rm BAO}=103.0~{\rm Mpc}.
\end{equation}
The same covariance matrix is used to generate and analyze the realizations, so the ensemble provides a direct closure test of the BAO likelihood and fitting pipeline.

The BAO MCMC campaign was run sequentially in the predetermined mock-index order. The analysis reported here comprises the first 192 independent BAO mock realizations, after which the campaign was terminated for computational economy. No realization within this sequence was chosen or omitted on the basis of its mock-data values or posterior results. These 192 completed BAO-only realizations constitute the recovery ensemble analyzed below. The initial fixed-length application of the chain-quality cuts
\begin{equation}
    N/\tau_{\max}>40,
    \qquad
    \hat R_{\max}<1.08,
\end{equation}
retained 132 realizations. We adaptively reran all 60 originally rejected realizations with their unchanged stored mock vectors and analysis setup; all ultimately satisfied the same criteria. The final BAO ensemble therefore contains 192/192 completed runs, a 100 per cent selection fraction. Its mean $N/\tau_{\max}$ is 53.0, its median is 46.3, its mean $\hat R_{\max}$ is 1.039, and its largest $\hat R_{\max}$ is 1.0797.

Table~\ref{tab:bao_ensemble_recovery} summarizes the BAO-only ensemble recovery statistics. The ensemble-mean posterior medians are
\begin{align}
    \langle\hat\Omega_{m0}\rangle
    &=0.2924\pm0.0023,
    \qquad
    \langle\hat w_0\rangle
    =-0.9503\pm0.0146,
    \nonumber\\
    \langle\hat w_a\rangle
    &=0.0162\pm0.0578,
    \qquad
    \langle\widehat{h r_d}\rangle
    =103.080\pm0.271~{\rm Mpc}.
\end{align}
Here, the quoted uncertainties are the standard errors of the ensemble-mean posterior medians rather than the posterior uncertainties of individual realizations. The corresponding mean central 68 per cent posterior half-widths are
\begin{align}
    \langle\sigma_{68}(\Omega_{m0})\rangle
    &=0.0545,
    \qquad
    \langle\sigma_{68}(w_0)\rangle
    =0.2490,
    \nonumber\\
    \langle\sigma_{68}(w_a)\rangle
    &=1.1023,
    \qquad
    \langle\sigma_{68}(h r_d)\rangle
    =4.1585~{\rm Mpc}.
\end{align}

For clarity, the uncertainty attached to an ensemble-mean median is determined by the realization-to-realization scatter of the medians. For example,
\begin{equation}
    {\rm SEM}(h r_d)
    =
    \frac{{\rm Scatter}(h r_d)}{\sqrt{N_{\rm sel}}}
    =
    \frac{3.756}{\sqrt{192}}
    \simeq
    0.271~{\rm Mpc}.
\end{equation}
By contrast,
\begin{equation}
    \left\langle\sigma_{68}(h r_d)\right\rangle
    =
    \frac{1}{N_{\rm sel}}
    \sum_{k=1}^{N_{\rm sel}}
    \frac{(h r_d)_{84}^{(k)}-(h r_d)_{16}^{(k)}}{2}
    =
    4.158~{\rm Mpc}
\end{equation}
is the mean central 68 per cent half-width of the individual marginalized posteriors. Thus, the relatively small uncertainty on $\langle\widehat{h r_d}\rangle$ does not mean that a single BAO-only realization determines $h r_d$ with a precision of $0.271~{\rm Mpc}$; the typical realization-level posterior uncertainty is approximately $4.2~{\rm Mpc}$.

\begin{table}[!htbp]
\centering
\caption{BAO-only ensemble recovery diagnostics for the $q_{\rm BAO}=h r_d$ parameterization. The table uses all 192 completed, converged BAO-only MCMC runs from the sequentially analyzed ensemble of independent mock realizations generated from the fiducial model. Here ``Scatter'' denotes the standard deviation of the posterior medians across realizations, while the uncertainty quoted on $\langle\hat\theta_{50}\rangle$ and on the ensemble-mean median offset is the standard error of the ensemble mean. Coverage is evaluated directly from $[p_{16},p_{84}]$ and $[p_{2.5},p_{97.5}]$. The BAO mock vectors were generated directly with the fiducial scale coordinate $q_{\rm BAO}^{\rm fid}=h r_d=103.0~{\rm Mpc}$.}
\label{tab:bao_ensemble_recovery}
\begin{tabular}{lccccccc}
\hline\hline
Parameter
& Fiducial
& $\langle\hat\theta_{50}\rangle$
& $\langle\hat\theta_{50}\rangle-\theta_{\rm fid}$
& Scatter
& $\langle\sigma_{68}\rangle$
& $C_{68}$
& $C_{95}$ \\
\hline
$\Omega_{m0}$
& 0.300
& $0.2924\pm0.0023$
& $-0.0076\pm0.0023$
& 0.0318
& 0.0545
& 82.81\%
& 100.00\% \\

$w_0$
& -1.000
& $-0.9503\pm0.0146$
& $0.0497\pm0.0146$
& 0.2029
& 0.2490
& 81.25\%
& 98.44\% \\

$w_a$
& 0.000
& $0.0162\pm0.0578$
& $0.0162\pm0.0578$
& 0.8013
& 1.1023
& 86.98\%
& 99.48\% \\

$h r_d$
& 103.000
& $103.080\pm0.271$
& $0.080\pm0.271$
& 3.756
& 4.158
& 73.96\%
& 97.92\% \\
\hline\hline
\end{tabular}
\end{table}

The realization-to-realization scatters of the posterior medians are
\begin{align}
    {\rm Scatter}(\Omega_{m0})&=0.0318,
    &
    {\rm Scatter}(w_0)&=0.2029,
    \nonumber\\
    {\rm Scatter}(w_a)&=0.8013,
    &
    {\rm Scatter}(h r_d)&=3.756~{\rm Mpc}.
\end{align}
The BAO posterior medians move substantially from one realization to another. For $w_a$, for example,
\begin{equation}
    \frac{{\rm Scatter}(w_a)}
    {\langle\sigma_{68}(w_a)\rangle}
    =
    \frac{0.8013}{1.1023}
    \simeq0.73.
\end{equation}
Thus, the elongated distribution of BAO posterior medians is produced by appreciable realization-to-realization motion along the weakly constrained BAO degeneracy direction, while each individual posterior remains broader still.

The marginal 68 per cent coverages are 82.81\%, 81.25\%, 86.98\%, and 73.96\% for $\Omega_{m0}$, $w_0$, $w_a$, and $h r_d$, respectively. The intervals are therefore generally conservative, although the $h r_d$ coverage is closer to the nominal 68 per cent target. The corresponding direct-percentile 95 per cent coverages are 100.00\%, 98.44\%, 99.48\%, and 97.92\%. The joint rectangular coverages are 77.08\% and 97.92\% in the $(w_0,w_a)$ sector and 67.19\% and 97.40\% in the full four-parameter space at 68 and 95 per cent, respectively. Thus, expanding the ensemble from the original 132 fixed-length-selected runs to all 192 converged runs leaves the key 95 per cent recovery conclusion essentially unchanged.

Figure~\ref{fig:bao_w0wa_scatter} provides a geometric view of this behavior. The posterior medians are not distributed isotropically around the fiducial point. Instead, they form an elongated point cloud aligned with the expected BAO degeneracy ridge in the $(w_0,w_a)$ plane. This orientation shows that changes in the mock BAO data move the marginalized median primarily along the weakly constrained direction. Consequently, the posterior median of an individual BAO realization can lie in a DDE-like region even though its broad posterior remains consistent with the fiducial model. A displaced median should therefore not, by itself, be interpreted as statistically significant evidence for DDE.

\begin{figure}[!htbp]
\centering
\includegraphics[width=0.74\linewidth]
{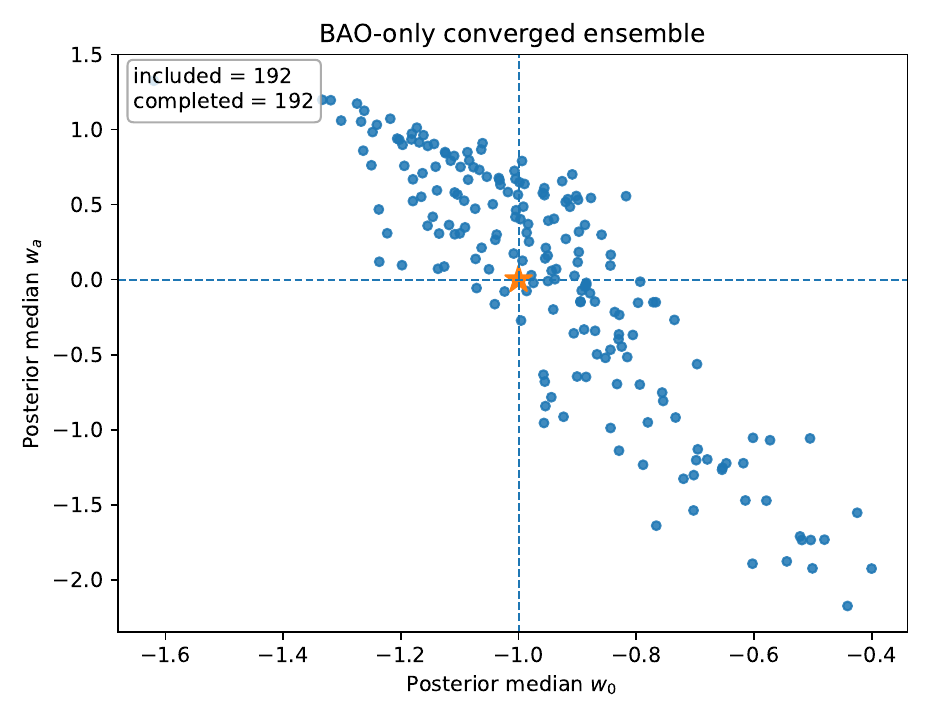}
\caption{BAO-only ensemble recovery in the $(w_0,w_a)$ plane for all 192 completed, converged realizations. Each point shows the posterior median of one mock realization, and the orange star marks the fiducial $\Lambda$CDM input $(w_0,w_a)=(-1,0)$. The point cloud is strongly elongated along the expected BAO degeneracy direction, showing that realization-to-realization motion occurs predominantly along the weakly constrained direction rather than across it. Consequently, a displaced posterior median need not imply statistically significant evidence for DDE.}
\label{fig:bao_w0wa_scatter}
\end{figure}

Figure~\ref{fig:bao_summary} summarizes the corresponding coverage and pull diagnostics. The mean realization-level pulls are
\begin{equation}
    -0.089,
    \qquad
    0.165,
    \qquad
    0.129,
    \qquad
    -0.028
\end{equation}
for $\Omega_{m0}$, $w_0$, $w_a$, and $h r_d$, respectively. Their realization-to-realization standard deviations are
\begin{equation}
    0.755,
    \qquad
    0.846,
    \qquad
    0.895,
    \qquad
    0.970.
\end{equation}
The pull for each realization is defined using the symmetrized central 68 per cent posterior width,
\begin{equation}
    p_\theta
    =
    \frac{\hat\theta_{50}-\theta_{\rm fid}}
    {(\theta_{84}-\theta_{16})/2}.
\end{equation}
The 68 and 95 per cent coverage indicators are evaluated directly using the marginalized percentile intervals $[\theta_{16},\theta_{84}]$ and $[\theta_{2.5},\theta_{97.5}]$, respectively. Joint rectangular coverage requires simultaneous inclusion of the relevant fiducial parameter values within their individual marginalized intervals.

The mean pulls are obtained by first computing the pull for each realization and then averaging over the full converged ensemble. They therefore need not equal the ensemble-mean median offset divided by the mean posterior width. The modest mean pulls, pull scatters of order unity, and high 95 per cent coverages show that the BAO-only median displacements are consistent with ordinary realization scatter along a broad degeneracy direction rather than a failure of fiducial-model recovery or a statistically significant false DDE signal.

\begin{figure}[!htbp]
\centering
\includegraphics[width=0.84\linewidth]
{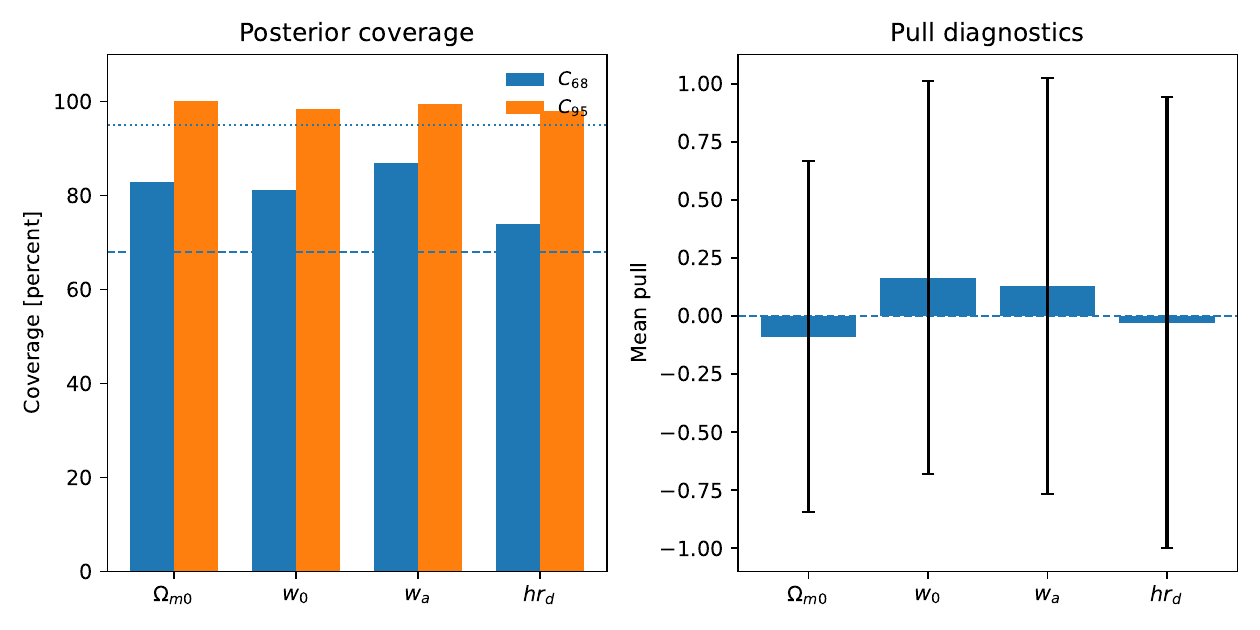}
\caption{Summary diagnostics for all 192 completed, converged BAO-only runs. Left: direct-percentile 68 and 95 per cent marginal posterior coverages for $\Omega_{m0}$, $w_0$, $w_a$, and $h r_d$; the dashed and dotted horizontal lines mark the corresponding nominal targets. Right: mean realization-level pulls for the same parameters, with error bars showing their realization-to-realization standard deviations. The generally conservative 68 per cent intervals, high 95 per cent coverages, and modest mean pulls show no evidence that the BAO-only likelihood spuriously excludes the fiducial input.}
\label{fig:bao_summary}
\end{figure}

\FloatBarrier

\subsection{CMB-only ensemble recovery}
\label{subsec:cmb_ensemble}

For the compressed CMB-only mock analysis, the sampler uses the internal parameter vector $(\omega_m,H_0,w_0,w_a,\omega_b)$, with $\omega_m=\Omega_{m0}h^2$, and transforms back to $\Omega_{m0}$ for reporting. This parameterization follows the physical matter-density combination to which the CMB distance prior is most directly sensitive. The likelihood is based on the three compressed observables $(R,\ell_A,\omega_b)$. The observable $\omega_b$ directly constrains the physical baryon density, whereas the two geometric quantities $R$ and $\ell_A$ constrain only limited combinations of $(\Omega_{m0},H_0,w_0,w_a)$. Consequently, the four-dimensional late-time sector remains strongly underconstrained and contains broad, asymmetric degeneracy directions. We therefore use posterior coverage and realization-level pull statistics as the primary recovery diagnostics in this CMB-only problem.

We completed 208 CMB-only mock realizations. With the baseline quality cuts
\begin{equation}
    N/\tau_{\max}>35,  \qquad \hat R_{\max}<1.09,
\end{equation}
101 realizations are retained. The selected baseline sample has median $N/\tau_{\max}=47.8$, median $\hat R_{\max}=1.063$, and a selection fraction of 48.6\%. As a stricter robustness check, we also impose
\begin{equation}
    N/\tau_{\max}>40, \qquad \hat R_{\max}<1.08,
\end{equation}
which retains 73 realizations. The strict sample has median $N/\tau_{\max}=50.5$, median $\hat R_{\max}=1.059$, and a selection fraction of 35.1\%. The relatively low retention fractions reflect the slow exploration of the broad CMB-only degeneracy directions. The agreement between the baseline and strict samples provides an internal check that the recovery conclusions are not sensitive to the precise quality threshold.

\subsubsection{CMB convergence-selection robustness}
\label{subsec:cmb_selection_audit}
The primary CMB classification uses the baseline $35/1.09$ criteria: 101 of 208 fixed-length runs are selected and 107 are rejected; the stricter $40/1.08$ split is only a robustness diagnostic. In representative 100,000-step extensions, mocks 2 and 5 reached $N/\tau_{\max}=145.0$ and 117.9 while walker-based split-$\hat R_{\max}$ remained at 1.100 and 1.153. Their walkers occupied a single connected degeneracy ridge, with the residual diagnostic controlled by $\omega_b$ and transformed $\Omega_{m0}$ rather than by $w_0$ or $w_a$.

We therefore tested selection dependence directly using all 208 original stored CMB mock vectors, without including any adaptive or smoke-test run. The selected and rejected mock-fluctuation amplitudes have means $\chi^2_{\rm mock}=3.049$ and 3.011, respectively, corresponding to a standardized effect $d=0.016$ with 95 per cent bootstrap interval $[-0.261,0.288]$ and permutation $p=0.909$. The local $w_0$ and $w_a$ response scores both give $d=-0.003$ ($p=0.984$); in this three-observable likelihood those two local response vectors are collinear, so this test constrains their shared CMB geometric-degeneracy direction rather than separating them. A ten-fold cross-validated classifier based on the three covariance eigenmodes plus $\chi^2_{\rm mock}$ gives ROC--AUC 0.420, consistent with its permutation null (two-sided $p=0.162$). Within the statistical sensitivity of this 208-mock ensemble, the baseline convergence classification is therefore not detectably associated with the amplitude or direction of the mock fluctuations relevant to the cosmological inference. This is not proof that the selection effect is exactly zero: individual effects of roughly 0.3--0.4 standardized units remain compatible with the finite sample. Appendix~\ref{app:cmb_selection_audit} gives the detailed audit.

Table~\ref{tab:cmb_ensemble_recovery} shows that the compressed CMB-only likelihood produces strongly displaced ensemble-mean posterior medians together with very broad marginalized intervals along the high-redshift distance-prior degeneracy. The ensemble-mean medians are
\begin{align}
    \langle\hat\Omega_{m0}\rangle &=0.2093\pm0.0002, \quad 
    \langle\hat H_0\rangle =83.97\pm0.03~ {\rm km\,s^{-1}\,Mpc^{-1}},
    \nonumber \\ 
    \langle\hat w_0\rangle &=-1.2370\pm0.0035, \quad 
    \langle\hat w_a\rangle =-0.7117\pm0.0039.
\end{align}
Here, the quoted uncertainties are the standard errors of the ensemble-mean posterior medians rather than the posterior uncertainties of individual realizations. The corresponding mean posterior widths are
\begin{align}
    \langle\sigma_{68}(\Omega_{m0})\rangle &=0.0749, \quad  \langle\sigma_{68}(H_0)\rangle =13.10~{\rm km\,s^{-1}\,Mpc^{-1}}, \nonumber 
    \\
    \langle\sigma_{68}(w_0)\rangle &=0.5240, \quad  \langle\sigma_{68}(w_a)\rangle =1.6137, \quad  \langle\sigma_{68}(\omega_b)\rangle =0.000146.
\end{align}

For clarity, the small uncertainties attached to the ensemble-mean posterior medians quantify the precision with which the mean locations of the realization-level medians are determined; they are not single-realization parameter constraints. For example,
\begin{align}
    {\rm SEM}(H_0)
    &=
    \frac{{\rm Scatter}(H_0)}{\sqrt{N_{\rm sel}}}
    =
    \frac{0.317}{\sqrt{101}}
    \simeq
    0.032~
    {\rm km\,s^{-1}\,Mpc^{-1}},
    \\
    {\rm SEM}(w_a)
    &=
    \frac{{\rm Scatter}(w_a)}{\sqrt{N_{\rm sel}}}
    =
    \frac{0.0397}{\sqrt{101}}
    \simeq
    0.0039.
\end{align}
By contrast, the corresponding mean central 68 per cent posterior half-widths are
\begin{align}
    \left\langle\sigma_{68}(H_0)\right\rangle
    &=
    \frac{1}{N_{\rm sel}}
    \sum_{k=1}^{N_{\rm sel}}
    \frac{H_{0,84}^{(k)}-H_{0,16}^{(k)}}{2}
    =
    13.10~
    {\rm km\,s^{-1}\,Mpc^{-1}},
    \\
    \left\langle\sigma_{68}(w_a)\right\rangle
    &=
    \frac{1}{N_{\rm sel}}
    \sum_{k=1}^{N_{\rm sel}}
    \frac{w_{a,84}^{(k)}-w_{a,16}^{(k)}}{2}
    =
    1.6137.
\end{align}
Thus, the compact realization-to-realization distribution of the posterior medians does not imply precise CMB-only constraints on $H_0$ or $w_a$. Instead, it shows that similarly displaced median locations recur across the ensemble even though the individual marginalized posteriors remain very broad along the CMB distance-prior degeneracy.

\begin{table}[!htbp]
\centering
\caption{CMB-only ensemble recovery diagnostics for the baseline quality cuts $N/\tau_{\max}>35$ and $\hat R_{\max}<1.09$. The baseline sample retains 101 out of 208 completed CMB-only mock realizations. Here ``Scatter'' denotes the standard deviation of the posterior medians across realizations, while the uncertainty quoted on $\langle\hat\theta_{50}\rangle$ and on the ensemble-mean median offset is the standard error of the ensemble mean. The broad and asymmetric late-time posterior intervals produce conservative coverage despite substantial displacements of their marginalized medians. Values involving $H_0$ are quoted in ${\rm km\,s^{-1}\,Mpc^{-1}}$.}
\label{tab:cmb_ensemble_recovery}
\begin{tabular}{lccccccc}
\hline\hline
Parameter
& Fiducial
& $\langle\hat\theta_{50}\rangle$
& $\langle\hat\theta_{50}\rangle-\theta_{\rm fid}$
& Scatter
& $\langle\sigma_{68}\rangle$
& $C_{68}$
& $C_{95}$ \\
\hline
$\Omega_{m0}$
& 0.3000
& $0.2093\pm0.0002$
& $-0.0907\pm0.0002$
& 0.0024
& 0.0749
& 100.0\%
& 100.0\% \\

$H_0$
& 70.00
& $83.97\pm0.03$
& $13.97\pm0.03$
& 0.32
& 13.10
& 99.0\%
& 100.0\% \\

$w_0$
& -1.0000
& $-1.2370\pm0.0035$
& $-0.2370\pm0.0035$
& 0.0352
& 0.5240
& 100.0\%
& 100.0\% \\

$w_a$
& 0.0000
& $-0.7117\pm0.0039$
& $-0.7117\pm0.0039$
& 0.0397
& 1.6137
& 100.0\%
& 100.0\% \\

$\omega_b$
& 0.022370
& $0.022334\pm0.000014$
& $-0.000036\pm0.000014$
& 0.000143
& 0.000146
& 67.3\%
& 97.0\% \\
\hline\hline
\end{tabular}
\end{table}

The realization-to-realization scatters of the posterior medians, especially
\begin{equation}
    {\rm Scatter}(w_0)=0.0352, \qquad  {\rm Scatter}(w_a)=0.0397,
\end{equation}
are much smaller than the corresponding marginalized posterior widths. Thus, the displaced median values recur in nearly the same region for many independent mock realizations. This behavior is consistent with a stable displacement produced by the nonlinear projection of the broad, asymmetric CMB-only posterior, with possible contributions from prior-volume effects, rather than being driven primarily by ordinary realization-to-realization noise. It should not be interpreted as a statistically significant preference for DDE.

The baseline 68 per cent coverages are 100.0\%, 99.0\%, 100.0\%, 100.0\%, and 67.3\% for $\Omega_{m0}$, $H_0$, $w_0$, $w_a$, and $\omega_b$, respectively. Thus, the marginalized intervals in the four-dimensional late-time sector are highly conservative, whereas the $\omega_b$ interval has nearly nominal 68 per cent coverage. The corresponding 95 per cent coverages are 100.0\%, 100.0\%, 100.0\%, 100.0\%, and 97.0\%. The joint rectangular 95 per cent coverage is 100.0\% in the $(w_0,w_a)$ sector and 97.0\% in the full five-parameter space. Accordingly, the displaced marginalized medians do not represent exclusion of the fiducial $\Lambda$CDM model.

Figure~\ref{fig:cmb_w0wa_scatter} provides a geometric diagnostic of the CMB-only posterior structure. The posterior medians are not broadly dispersed around the fiducial point; instead, they form a compact cloud near $(w_0,w_a)\simeq(-1.24,-0.71)$. The compactness of this cloud, compared with the much broader individual marginalized posteriors, shows that the displacement is not primarily caused by random realization-to-realization motion along the degeneracy direction. Rather, the repeated median location is consistent with the nonlinear projection of the broad CMB distance-prior degeneracy, with possible additional contributions from prior-volume effects. The figure visualizes a stable displacement of the marginalized median induced by the geometry of the underconstrained likelihood, not a statistically significant exclusion of the fiducial $(w_0,w_a)=(-1,0)$ point. Indeed, the fiducial values remain inside the reported 68 and 95 per cent intervals for every baseline realization in the dark-energy sector.

\begin{figure}[!htbp]
\centering
\includegraphics[width=0.74\linewidth]
{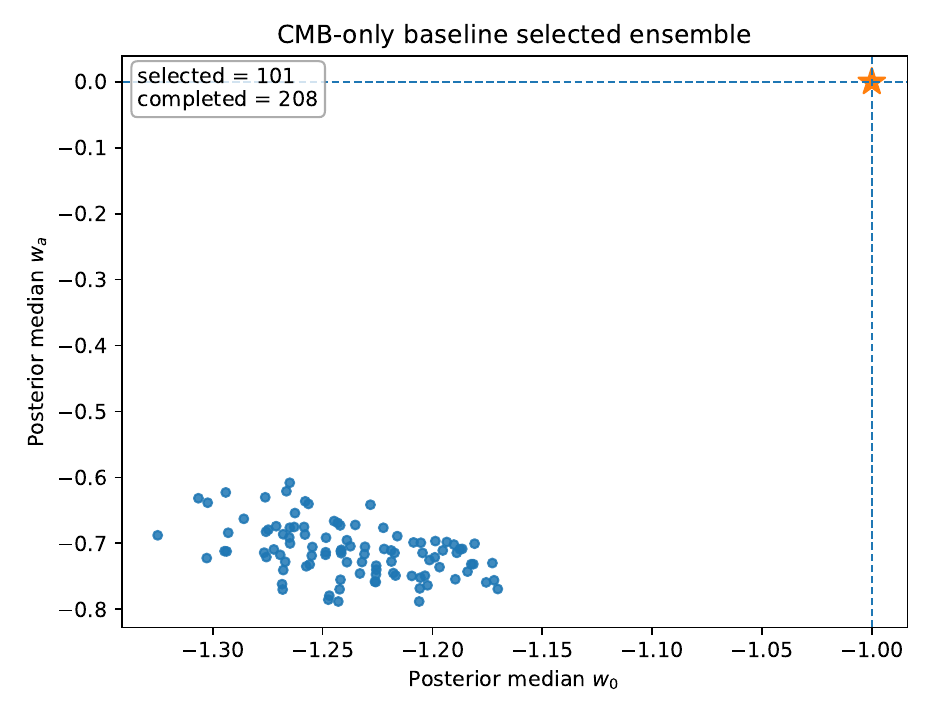}
\caption{CMB-only ensemble recovery in the $(w_0,w_a)$ plane for the 101 baseline quality-selected realizations. Each point shows the posterior median of one mock realization, and the orange star marks the fiducial $\Lambda$CDM input $(w_0,w_a)=(-1,0)$. The posterior medians form a compact cloud displaced from the fiducial point. Compared with the much broader individual posteriors, this compactness shows that the displacement is a stable consequence of the underconstrained marginalized-posterior geometry rather than evidence that the fiducial model is excluded.}
\label{fig:cmb_w0wa_scatter}
\end{figure}

Figure~\ref{fig:cmb_summary} complements the geometric picture in Fig.~\ref{fig:cmb_w0wa_scatter} by separating empirical interval coverage from median displacement. For the baseline sample, the mean realization-level pulls are
\begin{equation}
    -1.212,\qquad
    1.067,\qquad
    -0.452,\qquad
    -0.442,\qquad
    -0.247
\end{equation}
for $\Omega_{m0}$, $H_0$, $w_0$, $w_a$, and $\omega_b$, respectively. The larger and oppositely signed pulls in $\Omega_{m0}$ and $H_0$ reflect the characteristic CMB distance-prior degeneracy and its projection onto the marginalized one-dimensional posteriors, whereas the dark-energy pulls remain below about one half in magnitude.

The small realization-to-realization pull scatter for $\Omega_{m0}$, $H_0$, $w_0$, and $w_a$ shows that this pattern is repeated coherently across the ensemble. By contrast, the $\omega_b$ pull has a scatter close to unity and nearly nominal coverage, consistent with $\omega_b$ being directly constrained by the third compressed observable. The 95 per cent coverages remain at or above 97 per cent for all five reported parameters under both the baseline and strict quality cuts. Thus, the substantial shifts of the marginalized medians do not correspond to a failure of fiducial-model recovery.

The pull for each realization is defined using the symmetrized central 68 per cent posterior width,
\begin{equation}
    p_\theta = \frac{\hat\theta_{50}-\theta_{\rm fid}}{(\theta_{84}-\theta_{16})/2}.
\end{equation}
By contrast, the 68 and 95 per cent coverage indicators are evaluated directly using the marginalized percentile intervals $[\theta_{16},\theta_{84}]$ and $[\theta_{2.5},\theta_{97.5}]$, respectively. Because the CMB-only marginalized posteriors are strongly asymmetric, the pull based on a single symmetrized width need not correspond directly to percentile interval inclusion. A mean pull slightly larger than unity can coexist with high empirical coverage.

The robustness sample obtained with the stricter quality cuts gives mean dark-energy pulls
\begin{equation}
    \langle p_{w_0}\rangle=-0.452,
    \qquad
    \langle p_{w_a}\rangle=-0.445,
\end{equation}
and 95 per cent coverages of 100.0\%, 100.0\%, 100.0\%, 100.0\%, and 97.3\% for $\Omega_{m0}$, $H_0$, $w_0$, $w_a$, and $\omega_b$, respectively. The persistence of the same median displacements, pull pattern, and high coverages under the stricter diagnostic threshold, together with the direct mock-space audit above, supports the interpretation that the large marginalized-median shifts reflect the structure of the underconstrained likelihood rather than a detectable convergence-selection dependence, a statistically significant false DDE signal, or a spurious exclusion of the fiducial $\Lambda$CDM input.

\begin{figure}[!htbp]
\centering
\includegraphics[width=0.84\linewidth]
{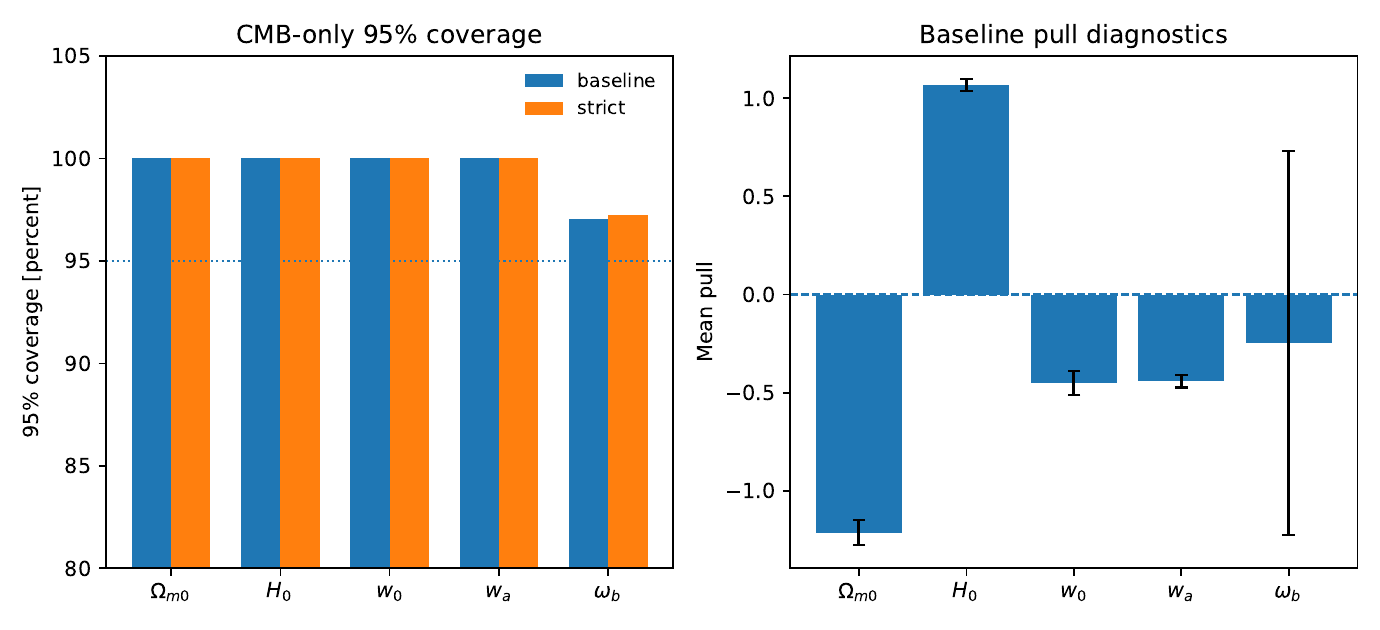}
\caption{Summary diagnostics for the CMB-only ensemble. Left: empirical 95 per cent marginal posterior coverages for the baseline and strict quality cuts. Right: baseline mean realization-level pulls, with error bars showing their realization-to-realization standard deviations. The dark-energy mean pulls remain below about one half in magnitude, whereas the larger and oppositely signed pulls in $\Omega_{m0}$ and $H_0$ reflect the broad CMB distance-prior degeneracy and asymmetric marginalization. The high 95 per cent coverages show that these median displacements do not correspond to spurious exclusion of the fiducial input.}
\label{fig:cmb_summary}
\end{figure}

\FloatBarrier

\subsection{SNe-only ensemble recovery}
\label{subsec:sne_ensemble}

The SNe-only ensemble uses the full Pantheon+-like covariance matrix and the same CPL background model as the other single-probe tests. Although the redshift sample and covariance are taken from the Pantheon+SH0ES release, the mock mean is the fiducial theoretical Hubble diagram; no SH0ES or Cepheid absolute-calibration information is imposed in this SNe-only likelihood. Therefore, the resulting block acts as an uncalibrated relative-distance likelihood.

In this case, the absolute normalization of the Hubble diagram is degenerate with the Hubble scale. We sample the coordinate
\begin{equation}
    \Delta M_{\rm eff} = \Delta M - 5\log_{10}\left(\frac{H_0}{H_{0,\rm ref}}\right),
\end{equation}
and transform back to the original $\Delta M$ coordinate for reporting. This is an exact reparameterization of the SNe-only likelihood, although the resulting marginalized posteriors remain dependent on the adopted sampling priors. The use of $\Delta M_{\rm eff}$ aligns the sampling coordinates with the normalization degeneracy and prevents the sampler from treating the SNe-only block as an independent absolute-scale measurement of $H_0$.

We completed 116 SNe-only MCMC analyses and applied the quality cuts
\begin{equation}
    N/\tau_{\max}>40, \qquad \hat R_{\max}<1.08.
\end{equation}
The initial fixed-length analysis retained 103 realizations. We adaptively reran all 13 originally rejected realizations using their unchanged stored mock vectors and analysis setup; every rerun ultimately satisfied the same criteria. The final SNe ensemble therefore contains 116/116 completed runs, a 100 per cent selection fraction, with median $N/\tau_{\max}=62.7$ and median $\hat R_{\max}=1.020$.

Table~\ref{tab:sne_ensemble_recovery} summarizes the SNe-only ensemble recovery statistics. The ensemble-mean posterior medians are
\begin{align}
    \langle\hat\Omega_{m0}\rangle &=0.3185\pm0.0051, \quad \langle\hat H_0\rangle =70.023\pm0.018~{\rm km\,s^{-1}\,Mpc^{-1}}, \nonumber\\
    \langle\hat w_0\rangle &=-0.9838\pm0.0135, \quad \langle\hat w_a\rangle =-0.4227\pm0.0643, \quad \langle\widehat{\Delta M}\rangle =0.0024\pm0.0009.
\end{align}
Here, the quoted uncertainties are the standard errors of the ensemble-mean posterior medians rather than the posterior uncertainties of individual realizations. The corresponding mean posterior widths are
\begin{align}
    \langle\sigma_{68}(\Omega_{m0})\rangle &=0.0893, \quad \langle\sigma_{68}(H_0)\rangle = 20.395~{\rm km\,s^{-1}\,Mpc^{-1}}, \nonumber \\
    \langle\sigma_{68}(w_0)\rangle &=0.1423, \quad \langle\sigma_{68}(w_a)\rangle =1.1964, \quad  \langle\sigma_{68}(\Delta M)\rangle =0.6514.
\end{align}

The uncertainty on $\langle\hat H_0\rangle$ is the standard error of the mean of the 116 realization-level posterior medians,
\begin{equation}
    {\rm SEM}(H_0)
    =
    \frac{{\rm Scatter}(H_0)}{\sqrt{N_{\rm sel}}}
    =
    \frac{0.198}{\sqrt{116}}
    \simeq 0.018,
\end{equation}
whereas
\begin{equation}
    \langle\sigma_{68}(H_0)\rangle
    =
    \frac{1}{N_{\rm sel}}
    \sum_{k=1}^{N_{\rm sel}}
    \frac{H_{0,84}^{(k)}-H_{0,16}^{(k)}}{2}
    =
    20.395~
    {\rm km\,s^{-1}\,Mpc^{-1}}
\end{equation}
is the mean width of the individual marginalized posteriors. Thus, the small ensemble standard error does not imply a precise SNe-only constraint on the absolute Hubble scale.

\begin{table}[!htbp]
\centering
\caption{SNe-only ensemble recovery diagnostics for the full-covariance Pantheon+-like mock likelihood. The table uses all 116 completed, converged SNe-only MCMC runs. The sampler uses the normalization-degenerate coordinate $\Delta M_{\rm eff}=\Delta M-5\log_{10}(H_0/H_{0,\rm ref})$ and transforms back to the original $\Delta M$ coordinate for reporting. Here ``Scatter'' denotes the standard deviation of the posterior medians across realizations, while the uncertainty quoted on $\langle\hat\theta_{50}\rangle$ and on the ensemble-mean median offset is the standard error of the ensemble mean. Values involving $H_0$ are quoted in ${\rm km\,s^{-1}\,Mpc^{-1}}$.}
\label{tab:sne_ensemble_recovery}
\begin{tabular}{lccccccc}
\hline\hline
Parameter
& Fiducial
& $\langle\hat\theta_{50}\rangle$
& $\langle\hat\theta_{50}\rangle-\theta_{\rm fid}$
& Scatter
& $\langle\sigma_{68}\rangle$
& $C_{68}$
& $C_{95}$ \\
\hline
$\Omega_{m0}$
& 0.300
& $0.3185\pm0.0051$
& $0.0185\pm0.0051$
& 0.0545
& 0.0893
& 91.4\%
& 100.0\% \\

$H_0$
& 70.0
& $70.023\pm0.018$
& $0.023\pm0.018$
& 0.198
& 20.395
& 100.0\%
& 100.0\% \\

$w_0$
& -1.000
& $-0.9838\pm0.0135$
& $0.0162\pm0.0135$
& 0.1451
& 0.1423
& 65.5\%
& 98.3\% \\

$w_a$
& 0.000
& $-0.4227\pm0.0643$
& $-0.4227\pm0.0643$
& 0.6927
& 1.1964
& 89.7\%
& 100.0\% \\

$\Delta M$
& 0.000
& $0.0024\pm0.0009$
& $0.0024\pm0.0009$
& 0.0099
& 0.6514
& 100.0\%
& 100.0\% \\
\hline\hline
\end{tabular}
\end{table}

The 68 per cent coverages are 91.4\%, 100.0\%, 65.5\%, 89.7\%, and 100.0\% for $\Omega_{m0}$, $H_0$, $w_0$, $w_a$, and $\Delta M$, respectively. Thus, the $w_0$ interval has coverage consistent with the nominal 68 per cent level for this ensemble size, whereas the intervals for $\Omega_{m0}$, $w_a$, $H_0$, and $\Delta M$ are conservative. In particular, the complete 68 per cent coverage of $H_0$ and $\Delta M$ reflects their very broad normalization-degenerate marginalized posteriors rather than direct absolute-scale information.

The corresponding 95 per cent coverages are 100.0\%, 100.0\%, 98.3\%, 100.0\%, and 100.0\%. The joint rectangular 95 per cent coverage is 98.3\% both in the $(w_0,w_a)$ sector and in the full reported five-parameter space. For completeness, the corresponding joint rectangular 68 per cent coverages are 63.8\% in the $(w_0,w_a)$ sector and 61.2\% in the full reported five-parameter space. These results show that including all 116 completed, converged runs leaves the SNe-only 95 per cent recovery conclusion unchanged.

Figure~\ref{fig:sne_w0wa_scatter} shows the distribution of SNe-only posterior medians in the $(w_0,w_a)$ plane. The point cloud is broad and asymmetric, reflecting the limited ability of relative-distance data alone to constrain the time-dependent CPL direction. The ensemble-mean displacement in $w_a$ is
\begin{equation}
    \langle\hat w_a\rangle-w_{a,\rm fid} = -0.4227,
\end{equation}
which is only about $0.36$ times the mean posterior width,
\begin{equation}
    \frac{|\langle\hat w_a\rangle-w_{a,\rm fid}|}{\langle\sigma_{68}(w_a)\rangle} \simeq 0.35.
\end{equation}
Together with the 89.7\% and 100.0\% coverages at the 68 and 95 per cent levels, this shows that the apparent $w_a$ offset is a manifestation of the broad SNe-only CPL degeneracy rather than significant evidence for time evolution.

\begin{figure}[!htbp]
\centering
\includegraphics[width=0.74\linewidth]
{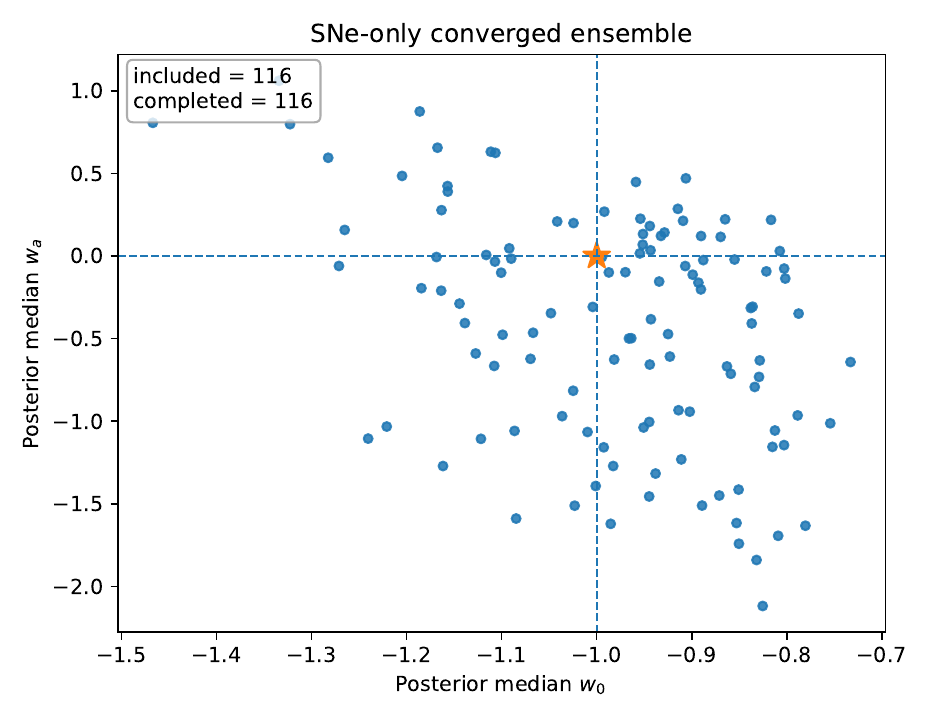}
\caption{SNe-only ensemble recovery in the $(w_0,w_a)$ plane for all 116 completed, converged realizations. Each point shows the posterior median of one mock realization, and the orange star marks the fiducial $\Lambda$CDM input $(w_0,w_a)=(-1,0)$. The broad and asymmetric point cloud reflects the limited ability of relative-distance data alone to constrain the time-dependent CPL direction. The ensemble-level displacement in $w_a$ remains small compared with the individual posterior widths and does not constitute significant evidence for DDE.}
\label{fig:sne_w0wa_scatter}
\end{figure}

The consequences of the normalization degeneracy for the marginalized posterior medians are illustrated in Fig.~\ref{fig:sne_h0_deltam_scatter}. The points cluster near the fiducial input in the $(H_0,\Delta M)$ plane, but this tight clustering must not be interpreted as a precise SNe-only determination of either parameter. The realization-to-realization scatters,
\begin{equation}
    {\rm Scatter}(H_0)=0.198~{\rm km\,s^{-1}\,Mpc^{-1}}, \qquad {\rm Scatter}(\Delta M)=0.0099,
\end{equation}
are much smaller than the corresponding individual posterior widths,
\begin{equation}
    \langle\sigma_{68}(H_0)\rangle = 20.395~{\rm km\,s^{-1}\,Mpc^{-1}}, \qquad \langle\sigma_{68}(\Delta M)\rangle = 0.6514.
\end{equation}
The tight clustering of the marginalized medians is consistent with the geometry of the normalization-degenerate posterior and with the stabilizing influence of the adopted parameterization and prior ranges, while the individual posteriors remain extremely broad along the degenerate direction. The SNe-only block consequently behaves as a relative-distance likelihood, as expected.

\begin{figure}[!htbp]
\centering
\includegraphics[width=0.74\linewidth]
{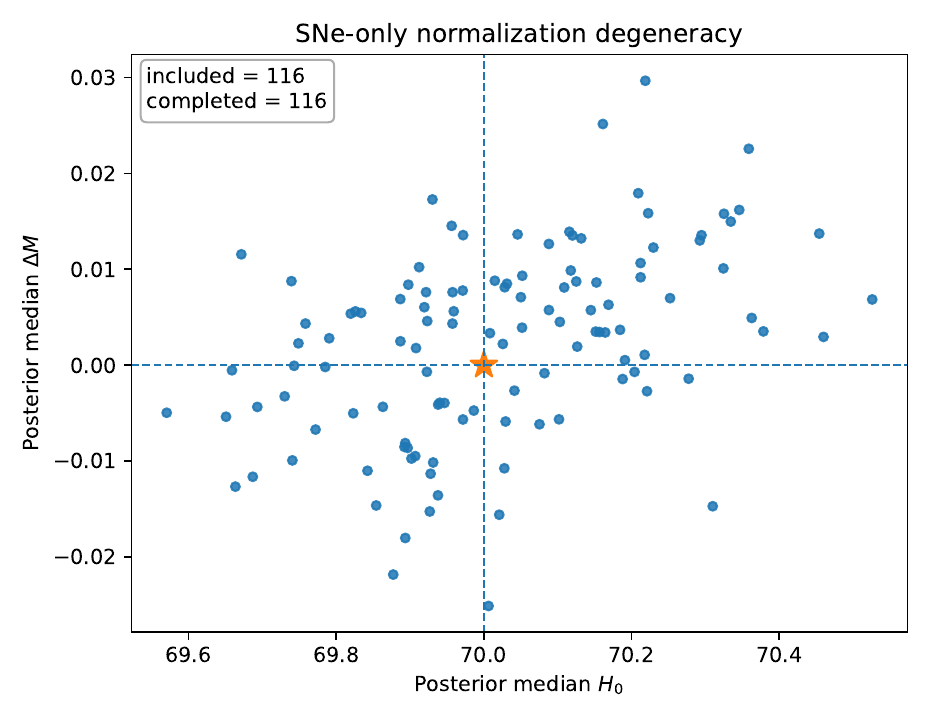}
\caption{SNe-only posterior medians in the $(H_0,\Delta M)$ plane for all 116 completed, converged realizations. The points show how the marginalized median locations vary across the ensemble. Their tight clustering near the fiducial input does not represent a precise SNe-only determination of either parameter, because the individual posteriors remain extremely broad along the normalization-degenerate direction.}
\label{fig:sne_h0_deltam_scatter}
\end{figure}

Figure~\ref{fig:sne_summary} summarizes the same behavior in coverage and pull space. The mean realization-level pulls are
\begin{equation}
    0.261,\qquad
    0.0011,\qquad
    0.326,\qquad
    -0.326,\qquad
    0.0037
\end{equation}
for $\Omega_{m0}$, $H_0$, $w_0$, $w_a$, and $\Delta M$, respectively. All mean pulls remain below one half in magnitude. The $w_0$ pull scatter is close to unity, consistent with its nearly nominal 68 per cent coverage, whereas the $w_a$ and $\Omega_{m0}$ intervals are more conservative.

The near-zero pulls and extremely small pull scatters for $H_0$ and $\Delta M$ do not indicate precise absolute-scale constraints. Rather, they arise because the realization-to-realization variation of their marginalized medians is negligible compared with their very large posterior widths. The high 95 per cent coverages and modest mean pulls therefore provide no evidence that the full-covariance SNe-only likelihood spuriously excludes the fiducial model or generates a statistically significant false DDE signal.

\begin{figure}[!htbp]
\centering
\includegraphics[width=0.84\linewidth]
{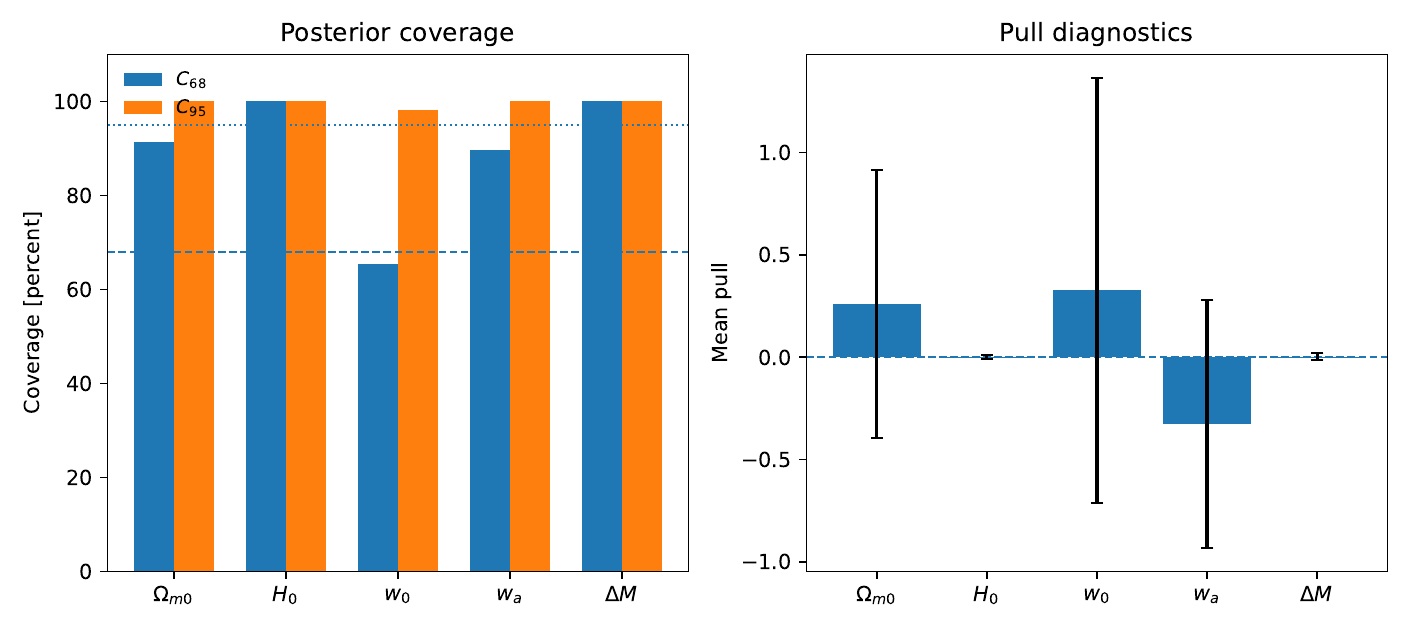}
\caption{Summary diagnostics for all 116 completed, converged SNe-only runs. Left: empirical 68 and 95 per cent marginal posterior coverages for $\Omega_{m0}$, $H_0$, $w_0$, $w_a$, and $\Delta M$. Right: mean realization-level pulls for the same parameters, with error bars showing their realization-to-realization standard deviations. The near-zero pulls and extremely small pull scatters for $H_0$ and $\Delta M$ reflect their very broad normalization-degenerate posteriors rather than precise absolute-scale constraints. The high 95 per cent coverages show that the full-covariance SNe-only likelihood does not spuriously exclude the fiducial input.}
\label{fig:sne_summary}
\end{figure}

\FloatBarrier

\subsection{Summary of ensemble recovery diagnostics}
\label{subsec:summary_table}

Table~\ref{tab:ensemble_summary} summarizes the ensemble validation results for the three single-probe likelihood blocks. The central conclusion is that none of the analyses spuriously excludes the fiducial $\Lambda$CDM input at the 95 per cent posterior-coverage level. The BAO-only and SNe-only rows include all completed runs after adaptive convergence; for the CMB-only analysis, the 101-run baseline selection is used as the main result, the stricter selection is only a threshold diagnostic, and the full 208-mock data-space audit finds no detectable selection dependence within its finite-sample sensitivity.

\begin{table}[!htbp]
\centering
\caption{Summary of ensemble recovery diagnostics. The columns $C_{95}^{\rm joint}(w_0,w_a)$ and $C_{95}^{\rm joint}({\rm all})$ denote rectangular direct-percentile 95 per cent coverage in the dark-energy subspace and in the full reported parameter space of each likelihood block, respectively. $N_{\rm comp}$ is the number of completed MCMC analyses, and $N_{\rm inc}$ is the number included in the reported statistics. All completed BAO and SNe runs are included after adaptive convergence. The CMB baseline row is the main CMB result, while the strict CMB row is only a threshold robustness diagnostic.}
\label{tab:ensemble_summary}
\begin{tabular}{lccccccc}
\hline\hline
Probe &
$N_{\rm comp}$ &
$N_{\rm inc}$ &
Quality cut &
$C_{95}(w_0)$ &
$C_{95}(w_a)$ &
$C_{95}^{\rm joint}(w_0,w_a)$ &
$C_{95}^{\rm joint}({\rm all})$ \\
\hline
BAO-only &
192 &
192 &
$N/\tau_{\max}>40,\ \hat R_{\max}<1.08$ &
98.44\% &
99.48\% &
97.92\% &
97.40\% \\
CMB-only &
208 &
101 &
$N/\tau_{\max}>35,\ \hat R_{\max}<1.09$ &
100.0\% &
100.0\% &
100.0\% &
97.0\% \\
CMB-only strict &
208 &
73 &
$N/\tau_{\max}>40,\ \hat R_{\max}<1.08$ &
100.0\% &
100.0\% &
100.0\% &
97.3\% \\
SNe-only &
116 &
116 &
$N/\tau_{\max}>40,\ \hat R_{\max}<1.08$ &
98.3\% &
100.0\% &
98.3\% &
98.3\% \\
\hline\hline
\end{tabular}
\end{table}

This table provides a compact numerical summary of Tables~\ref{tab:bao_ensemble_recovery}, \ref{tab:cmb_ensemble_recovery}, and \ref{tab:sne_ensemble_recovery}. The figures in the preceding subsections provide complementary visual diagnostics. The $(w_0,w_a)$ scatter plots show probe-dependent posterior-median displacements, while the coverage and pull panels show that these displacements do not correspond to significant failures of fiducial-model recovery within the individual likelihood blocks. In particular, the high 95 per cent coverages should be interpreted as evidence against spurious exclusion, rather than as proof that every marginalized interval is exactly calibrated at its nominal probability.

\FloatBarrier

\subsection{Implications for combined-probe constraints}
\label{subsec:combined_illustrative}

The three single-probe ensembles show that BAO, compressed CMB distance priors, and uncalibrated SNe exhibit distinct posterior geometries even when generated from the same fiducial $\Lambda$CDM cosmology. Because their dominant degeneracy directions are not aligned, their likelihood product is expected to reduce the allowed parameter volume substantially. This geometric complementarity does not, however, imply that the coverage of the product follows directly from the coverage of its constituent factors.

The present null tests therefore validate only the ensemble recovery of the three constituent likelihood blocks. A matched BAO+CMB+SNe ensemble would constitute a separate statistical experiment requiring explicit choices concerning realization matching, cross-probe covariance, joint quality selection, and multidimensional coverage. No claim is made here about the frequentist coverage, calibration, or false-exclusion rate of the full product likelihood.

\end{revblock}

\section{Conclusions}
\label{sec:conclusions}
\begin{revblock}

We have performed controlled ensemble null tests of BAO-like, compressed CMB-like, and SNe-like likelihood blocks generated from a common fiducial flat $\Lambda$CDM cosmology. The analysis quantifies recovery using percentile-based posterior coverage, realization-level pull statistics, and explicit convergence diagnostics. This ensemble formulation is necessary because a marginalized-posterior displacement in one mock realization cannot by itself establish either a systematic geometric bias or a physical preference for dynamical dark energy.

The main conclusion is that none of the three single-probe likelihood analyses spuriously excludes the fiducial flat $\Lambda$CDM input at the 95 per cent posterior-coverage level. All 192 completed BAO and all 116 completed SNe runs enter their final ensembles after adaptive convergence of the originally rejected chains. BAO-only recovery is assessed in the parameterization $(\Omega_{m0},w_0,w_a,h r_d)$, which respects the fact that BAO alone constrains the product $H_0r_d$ rather than $H_0$ and $r_d$ independently. The compressed CMB-only recovery analysis uses $\omega_m=\Omega_{m0}h^2$ internally; although the marginalized late-time posterior medians are strongly displaced along the high-redshift distance-prior degeneracy, the fiducial input remains inside the marginalized 95 per cent posterior intervals. Its baseline statistics use 101 of 208 fixed-length runs, while a direct audit of all 208 mock vectors finds no detectable dependence of that classification on the inference-relevant mock fluctuations within the finite ensemble's sensitivity. SNe-only recovery is assessed using $\Delta M_{\rm eff}$, which makes explicit the exact $H_0$--$\Delta M$ normalization degeneracy of an uncalibrated Hubble diagram.

The three likelihood blocks exhibit qualitatively different ensemble behavior. The BAO posterior medians move appreciably from realization to realization along the weakly constrained BAO degeneracy ridge. The compressed CMB-only medians instead recur in a compact displaced region, despite the much broader individual late-time posteriors, consistent with nonlinear marginalized-posterior geometry and possible prior-volume contributions. Representative long CMB extensions show that walker-based split-$\hat R$ can remain elevated even at very large $N/\tau$ because interacting walkers retain their relative locations along a connected ridge; it is therefore used here as a conservative heuristic rather than a calibrated independent-chain diagnostic. In the SNe-only analysis, the dark-energy parameters remain broadly constrained, while the tightly clustered $H_0$ and $\Delta M$ medians reflect the normalization-degenerate parameterization and very broad individual posteriors rather than a precise absolute-scale determination.

Therefore, the results do not support interpreting single-probe posterior shifts as stand-alone evidence for dynamical dark energy. Such shifts must be calibrated against mock ensembles before being assigned physical significance. In this sense, the null test provides a diagnostic warning: apparent deviations in compressed or single-probe likelihoods can arise from broad degeneracy directions, asymmetric marginalization, realization scatter, and prior-sensitive posterior geometry rather than from a genuine failure of $\Lambda$CDM.

This interpretation is consistent with the more general structural limitation of distance-based dark-energy inference. BAO, CMB distance priors, and SNe are primarily geometric probes: they constrain $w(z)$ only through the expansion history and through one or more integrals of that history. As shown in Ref.~\cite{Lee:2025jrr}, the mapping from perturbations in $w(z)$ to cosmological distances has a double-integral kernel structure, which acts as an intrinsic low-pass filter and leaves distance probes effectively sensitive to only a small number of smooth modes of the dark-energy equation of state. The present ensemble recovery tests should therefore not be read as showing that distance probes alone can fully characterize dynamical dark energy. Rather, they show that the geometric likelihood blocks considered here do not artificially exclude the fiducial $\Lambda$CDM model when analyzed with the appropriate calibration freedoms, parameterizations, percentile intervals, and convergence diagnostics.

A robust claim of time-varying dark energy must satisfy two separate requirements. First, it must survive ensemble coverage tests of the individual likelihood blocks and, where combined-probe claims are made, dedicated cross-probe consistency tests. Controlled tension-injection and forecasting analyses provide complementary tools for assessing how explicit inter-probe mismatches propagate into apparent EOS shifts~\cite{Lee:2025axp}. Second, a robust claim should be supported by observables with kernel structures complementary to distances. In particular, growth observables respond through differential dynamics and can activate additional Fisher-information directions that are structurally suppressed in distance-only analyses~\cite{Lee:2026sta}. Thus, the present work should be interpreted as a controlled validation of geometric single-probe likelihood blocks, not as a final test of the physical time dependence of dark energy.

The statistical scope of this work is the ensemble recovery and posterior geometry of the three constituent single-probe likelihood blocks. Correct recovery of these blocks does not by itself guarantee correct coverage of their full product, because the intersection of non-aligned likelihood ridges can amplify realization-dependent offsets. A matched combined-probe ensemble would constitute a separate statistical experiment requiring explicit choices concerning realization matching, cross-probe covariance, joint quality selection, and multidimensional coverage. No claim is therefore made about the frequentist coverage, calibration, or false-exclusion rate of the full BAO+CMB+SNe product.

For future analyses of DESI, Rubin/LSST, Euclid, and CMB data, the lesson is that claims of evolving $w(z)$ should be tested against ensemble recovery, explicit calibration freedoms, percentile coverage, and convergence diagnostics. A robust DDE claim should persist under these checks rather than appearing only as a displaced marginalized posterior in a single broad or partially constrained realization.

\end{revblock}

\appendix
\begin{revblock}
\section{Detailed compressed-CMB convergence-selection audit}
\label{app:cmb_selection_audit}
The audit uses the 208 original stored compressed-CMB mock vectors and their frozen fixed-length convergence classifications: 101 selected and 107 rejected under the primary $N_{\rm prod}/\tau_{\max}>35$, $\hat R_{\max}<1.09$ criteria. No adaptive extension or smoke-test run enters any test. The stricter $40/1.08$ classification is not used as the primary split.

For the overall mock fluctuation amplitude, the selected sample has $\chi^2_{\rm mock}$ mean/median/standard deviation $3.049/2.288/2.453$, compared with $3.011/2.549/2.310$ for the rejected sample. The standardized mean difference is $d=+0.016$, with 95 per cent bootstrap interval $[-0.261,+0.288]$, permutation $p=0.909$, and Kolmogorov--Smirnov $p=0.846$. The largest covariance-eigenmode effect is $d=-0.116$, with a max-$|d|$ permutation $p=0.790$. Across the partly redundant raw coordinates, the largest observed effect is the marginally standardized $\omega_b$ coordinate, $d=-0.154$ with interval $[-0.430,+0.118]$ and raw permutation $p=0.269$. There is no multiplicity-adjusted significant effect: the minimum Holm-adjusted $p$ for the predeclared $\chi^2$ plus eigenmode family is 1.0, and the minimum within-family Holm-adjusted $p$ over all tested features is 0.808.

A ten-fold cross-validated L2-regularized logistic regression using the three covariance eigenmodes and $\chi^2_{\rm mock}$ yields ROC--AUC 0.420; repeated cross-validation gives $0.428\pm0.029$. The label-permutation null has mean 0.490 and 95 per cent interval $[0.361,0.588]$, with two-sided $p=0.162$. Thus the observed below-0.5 value is not evidence for a reproducible inverse classifier and supplies no useful predictive signal beyond the permutation null.

The fluctuation amplitude is not detectably correlated with either convergence diagnostic: Spearman $\rho=+0.095$ ($p=0.177$) for $N_{\rm prod}/\tau_{\max}$ and $\rho=+0.070$ ($p=0.325$) for $\hat R_{\max}$. The largest $N/\tau$ association is with standardized $R$, $\rho=+0.132$ ($p=0.0536$), and the largest split-$\hat R$ association is with standardized $\omega_b$, $\rho=+0.110$ ($p=0.114$). The minimum adjusted correlation $p$ is 0.804.

Finally, the local cosmological-response projections give $d=-0.003$ and $p=0.984$ for both the $w_0$ and $w_a$ scores. Their equality reflects the collinearity of the two local response vectors in this three-observable compressed likelihood: the audit tests their shared geometric-degeneracy direction rather than resolving the two parameters independently. The largest physical-direction effect is the $\omega_b$ score, $d=-0.058$ with interval $[-0.332,+0.214]$. The largest absolute endpoints of the individual 95 per cent effect-size intervals are about 0.3--0.4 standardized units. These are finite-sample sensitivity bounds, not evidence that effects of those sizes exist. The supported conclusion is therefore that no selection dependence is detected within the sensitivity of the 208-mock ensemble, not that selection bias is proven to be exactly zero.
\end{revblock}

\section*{Declaration of generative AI and AI-assisted technologies}
\begin{revblock}
During the preparation of this work, the author used ChatGPT (OpenAI) for English-language editing and to assist in interpreting error messages encountered during numerical-code execution. The scientific content, equations, numerical methods, code development, analysis choices, and validation of the results were carried out and critically reviewed by the author. All AI-assisted suggestions were independently assessed by the author, who takes full responsibility for the content, results, and conclusions of this work.
\end{revblock}

\end{document}